\documentclass{scrartcl}

\usepackage[utf8]{inputenc}
\usepackage[T1]{fontenc}
\usepackage[UKenglish]{babel}
\usepackage{amsmath}
\usepackage{amssymb}
\usepackage{amsfonts}
\usepackage{nicefrac}
\usepackage{tikz}
\usepackage{subcaption}
\usepackage{booktabs}
\usepackage{authblk}
\usepackage{hyperref}

\usepackage[paper=a4paper,left=31mm,right=31mm,top=30mm, bottom=25mm]{geometry}

\DeclareOldFontCommand{\bf}{\normalfont\bfseries}{\mathbf}
\DeclareOldFontCommand{\it}{\normalfont\itshape}{\mathit}

\newtheorem{theorem}{Theorem}
\newtheorem{proposition}[theorem]{Proposition}
\newtheorem{lemma}[theorem]{Lemma}
\newtheorem{corollary}[theorem]{Corollary}
\newtheorem{example}{\it Example}
\newtheorem{definition}{Definition}

\newenvironment{proof}{\noindent{\bf Proof~}}{\null\hfill $\Box$\par\medskip}

\newcommand{\tws} {\text{tw}}
\newcommand{\pws} {\text{pw}}
\newcommand{\bigo}{\ensuremath{\mathcal{O}}}
\newcommand{\dist}{\text{dist}}
\newcommand{\co} {\text{co-}}
\newcommand{\N}{\mathbb{N}}
\newcommand{\calX}{\mathcal{X}}
\newcommand{\calZ}{\mathcal{Z}}
\newcommand{\calY}{\mathcal{Y}}
\newcommand{\calG}{\mathcal{G}}
\newcommand{\catprod}{\star_{\text{cat}}}
\newcommand{\ident}{\textrm{Ident}}
\newcommand{\subdiv}{\textrm{Subdiv}}

\setkomafont{author}{\sffamily}

\begin{document}
\def \gnode {node[circle, draw, inner sep=0pt, minimum width=3ex]}
\def \bag {node[circle, draw, inner sep=0pt, minimum width=3ex, font=\tiny]}

\title{The Behavior of Tree-Width and Path-Width under Graph Operations and Graph Transformations}

\author[1,*]{Frank Gurski}

\author[1]{Robin Weishaupt}

\affil[1]{\small University of  D\"usseldorf,
Institute of Computer Science, 
40225 D\"usseldorf, Germany}

\affil[*]{Corresponding Author: \href{mailto:frank.gurski@hhu.de}{frank.gurski@hhu.de}}

\date{\vspace{-5ex}}

\maketitle

\begin{abstract}
  Tree-width and path-width are well-known graph parameters.
  Many NP-hard graph problems allow polynomial-time solutions, when
  restricted to graphs of bounded tree-width or bounded path-width.
  In this work, we study the behavior of tree-width and path-width
  under various unary and binary graph transformations.
  Doing so, for considered transformations we provide upper and lower
  bounds for the tree-width and path-width of the resulting graph in
  terms of the tree-width and path-width of the initial graphs or argue
  why such bounds are impossible to specify.
  Among the studied, unary transformations are vertex addition, vertex
  deletion, edge addition, edge deletion, subgraphs, vertex
  identification, edge contraction, edge subdivision, minors, powers of
  graphs, line graphs, edge complements, local complements, Seidel
  switching, and Seidel complementation.
  Among the studied, binary transformations we consider the disjoint
  union, join, union, substitution, graph product, 1-sum, and corona of
  two graphs.

\end{abstract}

%%%%%%%%%%%%%%%%%%%%%%%%%%%%%%%%%%%%%%%%%%%%%%%%%%%%%%%%%%%%%%%%%%%%%%%%%%%
%%%%%%%%%%%%%%%%%%%%%%%%%%%%%%%%%%%%%%%%%%%%%%%%%%%%%%%%%%%%%%%%%%%%%%%%%%%
%%%%%%%%%%%%%%%%%%%%%%%%%%%%%%%%%%%%%%%%%%%%%%%%%%%%%%%%%%%%%%%%%%%%%%%%%%%
\section{Introduction}\label{sec-intro}

A {\em graph parameter} is a function that associates every graph with
a non-negative integer.
One of the most famous graph parameters is tree-width, which was
defined by Robertson and Seymour~\cite{RS86}.
Graphs of bounded tree-width are interesting from an algorithmic point of
view, as several NP-complete graph problems can be solved in
polynomial time for graph classes of bounded tree-width.
For example, tree-decompositions allow for many efficient algorithms
in dynamic programming ~\cite{Arn85,AP89,Hag00,KZN00}.
The same holds for the similar graph parameter path-width.
That is, because every path-decomposition can be interpreted as a
special case of a tree-decomposition.
Both parameters play a crucial role in the field of structural graph
theory, especially in the graph minor theory of Robertson and
Seymour~\cite{RS85}.

Trees and forests have tree-width at most one.
Series parallel graphs have tree-width at most two~\cite{Bod98}.
Outerplanar graphs (and subclasses, such as cactus graphs and maximal
outerplanar graphs) have tree-width at most two and $k$-outerplanar
graphs have tree-width at most $3k-1$~\cite{Bod98}.
Halin graphs have tree-width at most three~\cite{Bod98}.
For further classes of graphs with bounded tree-width, we refer to the
works by Bodlaender~\cite{Bod86,Bod88b,Bod98}.

Determining whether the tree-width or path-width of a graph
is at most a given value $w$ is NP-complete~\cite{ACP87}.
However, for every fixed integer $k$, one can decide in linear time
whether a given graph $G$ has tree-width or path-width $k$, see
Bodlaender~\cite{Bod96}.
For an in-depth overview of tree-width and path-width, we refer again
to the work by Bodlaender~\cite{Bod98}.

A {\em graph transformation} $f$ is a function that creates a new
graph $f(G_1,\ldots,G_n)$ from a number of $n \geq 1$ input graphs
$G_1, \ldots, G_n$.
Examples of graph transformations are taking an induced subgraph of a
graph, adding an edge to a graph, or generating the join of two
graphs.
A {\em graph operation} is a graph transformation which is
deterministic and invariant under isomorphism.
Examples of graph operations are the edge complementation of a graph
or generating the join of two graphs.\footnote{Please note that by
our definition the two graph transformations taking an induced
subgraph of a graph and adding an edge to a graph are no graph
operations.}
The graph theory books by Bondy and Murty~\cite{BM76} and
Harary~\cite{Har69} provide a large number of graph transformations.

The impact of graph operations, which can be defined by monadic second
order formulas (so-called MS transductions), on graph parameters can
often be shown in a very short way.
Unfortunately, the resulting bounds are most of the time rather
imprecise~\cite{Cou06,CE12}.

Transformations that reduce graphs can be used to characterize classes
of graphs by forbidden subgraphs.
For example, the property that a graph has tree-width at most $k$ is
preserved under the graph transformation ``taking minors''.
This fact is used to show that the set of graphs of tree-width at most
$k$ can be characterized by a finite set of forbidden
minors~\cite{RS85}.

The effect of graph transformations on graph parameters is
well studied, e.g., for band-width~\cite{CO86}, for
tree-width~\cite{Bod98}, for clique-width~\cite{Cou14,HOSG08,Gur17},
and for rank-width~\cite{HOSG08}.

In this paper we study the behavior of tree-width and path-width
under various graph transformations and graph operations.
Thereby, we consolidate known results from various works and prove
novel results.
Doing so, this work tries to provide a comprehensive overview of the
effects of unary and binary graph transformations on tree-width and
path-width.
Therefore, the paper is organized as follows.
In Section~\ref{intro}, we recall the definitions of tree-width and
path-width.
In Section~\ref{sec-un}, we consider the effects of the unary graph
transformations
vertex deletion,
vertex addition,
edge deletion,
edge addition,
subgraphs,
vertex identification,
edge contraction,
edge subdivision,
minors,
powers of graphs,
line graphs,
edge complements,
local complements,
Seidel switching,
and Seidel complementation on tree-width and path-width.
Whenever it is possible to bound the tree-width or path-width of the
resulting graph $f(G)$, we show how to compute the corresponding
decomposition in time linear in the size of the corresponding
decomposition for $G$.
In Section~\ref{section-binary-operations}, we give an overview of the
effects of the binary graph operations
disjoint union,
join,
union,
substitution,
graph products,
1-sum,
and corona on tree-width and path-width.
If it is possible to bound the tree-width or path-width of the
combined graph $f(G_1,G_2)$ in the tree-width or path-width of graphs
$G_1$ and $G_2$, we show how to compute the corresponding
decomposition in time linear in the size of the corresponding
decompositions for $G_1$ and $G_2$.
Finally, we summarize our results and provide some conclusions as well
as an outlook for future work in Section~\ref{sec-con}.

%%%%%%%%%%%%%%%%%%%%%%%%%%%%%%%%%%%%%%%%%%%%%%%%%%%%%%%%%%%%%%%%%%%%%%%%%%%
%%%%%%%%%%%%%%%%%%%%%%%%%%%%%%%%%%%%%%%%%%%%%%%%%%%%%%%%%%%%%%%%%%%%%%%%%%%
%%%%%%%%%%%%%%%%%%%%%%%%%%%%%%%%%%%%%%%%%%%%%%%%%%%%%%%%%%%%%%%%%%%%%%%%%%%
\section{Preliminaries}\label{intro}

%%%%%%%%%%%%%%%%%%%%%%%%%%%%%%%%%%%%%%%%%%%%%%%%%%%%%%%%%%%%%%%%%%%%%%%%%%%
\subsection{Graphs}
We work with finite, undirected {\em graphs} $G$, where $V(G)$
is a finite set of {\em vertices} and $E(G) \subseteq \{ \{u,v\} \mid
u,v \in V(G),~u \not= v\}$ is a finite set of {\em
edges} without loops or multiple edges.
For a vertex $v \in V(G)$, we denote by $N_G(v)$ the set of all vertices
adjacent to $v$ in $G$, i.e., $N_G(v) = \{ w\in V(G) \mid \{v,w\} \in
E(G) \}$.
The vertex set $N_G(v)$ is called the set of all {\em neighbors} of $v$ in
$G$ or the {\em neighborhood} of $v$ in $G$.
Note that $v$ itself does not belong to $N_G(v)$.
The {\em degree} of a vertex $v \in V(G)$, denoted by $\deg_G(v)$, is
the number of neighbors of $v$ in $G$, i.e., $\deg_G(v) = |N_G(v)|$,
and the {\em maximum degree} of $G$, denoted by $\Delta(G)$, is the
maximum over the vertices' degrees, i.e., $\Delta(G) = \max_{v \in V(G)}
\deg_G(v)$.
In this work, we discuss graphs only up to isomorphism.
This allows us to define the path with $n$ vertices as $P_n = (\{ v_1,
\ldots, v_n\}, \{ \{ v_1, v_2 \}, \ldots, \{ v_{n-1}, v_n \} \})$,
which is useful in several examples.
For definitions of further special graphs, we refer to the book by
Brandst\"adt et al.~\cite{BLS99}.

%%%%%%%%%%%%%%%%%%%%%%%%%%%%%%%%%%%%%%%%%%%%%%%%%%%%%%%%%%%%%%%%%%%%%%%%%%%
\subsection{Tree-width}\label{sec-tw}
One of the most famous tree structured graph classes are graphs of
bounded tree-width.
The first notion equivalent to tree-width, the dimension of a graph,
was introduced by Bertel\`{e} and Brioschi~\cite{BB73} in 1973.
The subsequent definition of tree-width was given by Robertson and
Seymour~\cite{RS86} in 1986.

\begin{definition}[Tree-width]
  \label{DTW}
  A {\em tree-decomposition} of a graph $G$ is a pair
  $(\calX, T)$, where $T$ is a tree and $\calX = \{ X_{u}
  \subseteq V(G) \mid u \in V(T) \}$ is a family of subsets of $V(G)$,
  called {\em bags}, such that the following three conditions hold:

  \begin{description}
  \item[{\textnormal{(tw-1)}}]
    $\bigcup_{u \in V(T)} X_u = V(G)$,\footnote{Please note that
      (tw-1) has no influence on the width of a tree-decomposition. It
      merely ensures that isolated vertices are covered by a
      tree-decomposition, which is useful for the design of algorithms along
      tree-decompositions.\label{fn1}}
  \item[{\textnormal{(tw-2)}}]
    for every edge $\{v_1, v_2\} \in E(G)$, there is a node $u \in V(T)$ with
    $v_1, v_2 \in X_u$, and
  \item[{\textnormal{(tw-3)}}]
    for every vertex $v \in V(G)$ the subgraph of $T$ induced by the nodes $u \in
    V(T)$ with $v \in X_u$ is connected.
  \end{description}
  The {\em width} of a tree-decomposition $( \calX = \{ X_u \mid u \in
  V(T) \}, T)$ is defined as
  \begin{align*}
    \max_{u \in V(T)} |X_u| - 1.
  \end{align*}
  Finally, the {\em tree-width} of a graph $G$, denoted by $\tws(G)$, is the
  smallest integer $k$, such that there is a tree-decomposition
  $(\calX, T)$ for $G$ of width $k$.
\end{definition}

Obviously, the tree-width of every graph $G$ is bounded by $|V(G)| -
1$, as one can always place all its vertices in a single bag.
Contrarily, determining whether the tree-width of some given graph is
at most some given integer $w$ is NP-complete, even for bipartite
graphs and complements of bipartite graphs~\cite{ACP87}.

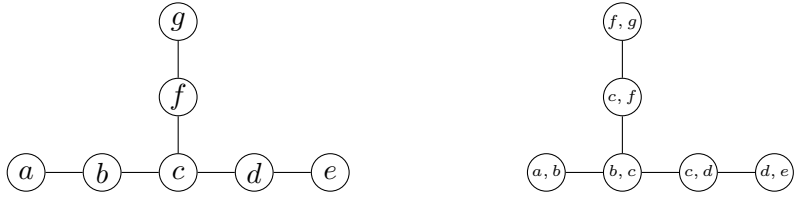
\begin{figure}
  \centering
  \begin{subfigure}{0.40\textwidth}
    \centering
    \begin{tikzpicture}
      \draw (-2,0) \gnode (a) {$a$};
      \draw (-1,0) \gnode (b) {$b$};
      \draw (0,0) \gnode (c) {$c$};
      \draw (1,0) \gnode (d) {$d$};
      \draw (2,0) \gnode (e) {$e$};
      \draw (0,1) \gnode (f) {$f$};
      \draw (0,2) \gnode (g) {$g$};
      \draw[-] (a) to (b);
      \draw[-] (b) to (c);
      \draw[-] (c) to (f);
      \draw[-] (f) to (g);
      \draw[-] (c) to (d);
      \draw[-] (d) to (e);
    \end{tikzpicture}
  \end{subfigure}
  \hspace{1ex}
  \begin{subfigure}{0.40\textwidth}
    \centering
    \begin{tikzpicture}
      \draw (-1,0) \bag (ab) {$a, b$};
      \draw (0,0) \bag (bc) {$b, c$};
      \draw (0,1) \bag (cf) {$c, f$};
      \draw (0,2) \bag (fg) {$f, g$};
      \draw (1,0) \bag (cd) {$c, d$};
      \draw (2,0) \bag (de) {$d, e$};
      \draw[-] (ab) to (bc);
      \draw[-] (bc) to (cf);
      \draw[-] (cf) to (fg);
      \draw[-] (bc) to (cd);
      \draw[-] (cd) to (de);
    \end{tikzpicture}
  \end{subfigure}
  \caption{A graph (left) and an associated tree-decomposition (right).}
  \label{F-ex-tw}
\end{figure}

\begin{example}[Tree-decomposition]
  In Figure~\ref{F-ex-tw} we depict a graph together with a corresponding
  tree-decomposition of width $1$.
\end{example}

Next to the previous definition of tree-width, tree-width can also be
defined by bramble numbers~\cite{Ree97} (also known as the size of
screens~\cite{ST93}) or partial $k$-trees~\cite{Ros74}.
We refer to the work by Bodlaender~\cite{Bod98}
for an overview of equivalent definitions.

Throughout the course of our work, we make use of the following lemma
by Scheffler~\cite{Sch89} several times.

\begin{lemma}\label{le-tw-deg-v}
  Every graph of tree-width at most $k$ has a vertex of degree at most
  $k$.
\end{lemma}

\begin{proof}
  Let $G$ be a graph with $\tws(G) = k$ and $(\calX, T)$ a
  tree-decomposition of width $k$ for $G$, such that $|V(T)|$ is as
  small as possible.
  Let $\ell$ be a leaf of $T$ and denote its parent in $T$ as $v$.
  Since $|V(T)|$ is as small as possible, there must exist a vertex $u
  \in X_\ell$ with $u \notin X_v$.
  Otherwise, $X_{\ell} \subset X_v$ holds and we can delete
  $X_{\ell}$, contradicting the minimality of $V(T)$.
  Thus, due to (tw-2) and (tw-3), all neighbors of $u$ must be in
  $X_\ell$ and, as $X_\ell$ contains at most $k+1$ vertices, it follows
  that $u$ has degree at most $k$.
\end{proof}

The subsequent proposition is stated by Bodlaender~\cite{Bod98}, who
in turn credits it to Rose~\cite{Ros74}.
As Rose's proof makes use of $k$-trees, subsequently, for the sake of
self-containment, we provide a proof only making use of
Lemma~\ref{le-tw-deg-v} and tree-width.

\begin{proposition}\label{prop-tw-edges}
  Let $G$ be a graph.
  Then, it holds that
  \begin{align*}
    |E(G)| \leq \tws(G) |V(G)| - \frac{1}{2} \tws(G) (\tws(G) + 1).
  \end{align*}
\end{proposition}

\begin{proof}
  By Lemma~\ref{le-tw-deg-v} we know that there exists at least one
  vertex $v$ in $G$ of degree at most $\tws(G)$, such that $v$
  possesses at most $\tws(G)$ incident edges in $G$.
  Let us denote the graph we obtain by removing $v$ from $G$ as
  $G_1$.
  Then, we know that from $G$ to $G_1$ we removed at most
  $\tws(G)$ edges from $G$.
  It is easy to see that removing $v$ and all its incident edges
  from $G$ can reduce the tree-width of $G$ but not increase it.
  Hence, $\tws(G_1) \leq \tws(G)$ holds and we can repeat the previous
  argument for $G_1$, i.e., remove a vertex as well as at most
  $\tws(G)$ edges from $G_1$.
  Repeating this argument for all vertices of $G$ results in $|E(G)|
  \leq \tws(G) |V(G)|$.
  However, at one point, there are at most $\tws(G)$ vertices left.
  This remaining graph can possess a tree-width of at most
  $\tws(G) - 1$, such that by Lemma~\ref{le-tw-deg-v} it follows that
  in this graph there is a vertex with degree at most $\tws(G) - 1$.
  Removing this vertex from the remaining graph, we remove at most
  $\tws(G) - 1$ edges.
  Repeating this argument for all of the $\tws(G)$ remaining vertices
  in sequence yields
  \begin{align*}
    |E(G)| & \leq \tws(G) (|V(G)| - \tws(G)) + \sum_{i = 1}^{\tws(G)} (\tws(G)
    - i) \\
    & = \tws(G) (|V(G)| - \tws(G)) + \tws(G)\tws(G) - \sum_{i=1}^{\tws(G)} i \\
    & = \tws(G) |V(G)| - \frac{1}{2} \tws(G) (\tws(G) + 1).
  \end{align*}
\end{proof}

While the previous proposition provides an upper bound for the number
of edges of a graph in terms of the number of its vertices as well as
its tree-width, the ensuing proposition by Kneis et al.~\cite{KMRR09}
provides an upper bound for the tree-width of a graph in terms of the
number of its edges and vertices.

\begin{proposition}
  Let $G$ be a graph.
  Then, it holds that
  \begin{align*}
    \tws(G)\leq \frac{|E(G)|}{5.769} + \bigo(\log(|V(G)|)).
  \end{align*}
\end{proposition}

Recall that a graph is said to be $k$-connected if and only if every
pair of its vertices is connected by at least $k$ vertex-disjoint
paths~\cite{Whi32}.
Making use of Lemma~\ref{le-tw-deg-v}, we obtain the following lower
bound for the tree-width of $k$-connected graphs.

\begin{lemma}\label{le-tw-con}
  Every $k$-connected graph has tree-width at least $k$.
\end{lemma}

\begin{proof}
  Let $G$ be a $k$-connected graph.
  By definition, it holds that for all vertices $u,v \in V(G)$ there
  exists $k$ vertex-disjoint paths from $u$ to $v$ in $G$.
  Consequently, $\deg_G(u) \geq k$ as well as $\deg_G(v) \geq k$ must
  be true.
  In other words, every vertex of $G$ has degree at least $k$.
  By the contraposition of Lemma~\ref{le-tw-deg-v} we immediately
  obtain $\tws(G) \geq k$.
\end{proof}

To conclude the preliminary subsection on tree-width, we cite two
lemmas by Bodlaender and M\"ohring~\cite{BM93} of which we make use in
future sections.

\begin{lemma}[Clique containment]\label{le-tw-clique}
  Let $(\calX, T)$ be a tree-decomposition for a graph $G$ and $C
  \subseteq V(G)$ a clique in $G$.
  Then, there exists some $u \in V(T)$ with $C \subseteq X_u$.
\end{lemma}

\begin{lemma}[Complete bipartite subgraph
  containment]\label{le-tw-bip-clique}
  Let $(\calX, T)$ be a tree-decomposition for a graph $G$ and $A, B
  \subseteq V(G)$ with $A \cap B = \emptyset$ and $\{ \{ u, v \} \mid u
  \in A, v \in B \} \subseteq E(G)$.
  Then, there exists a vertex $u \in V(T)$ with $A \subseteq X_u$ or
  $B \subseteq X_u$.
\end{lemma}

%%%%%%%%%%%%%%%%%%%%%%%%%%%%%%%%%%%%%%%%%%%%%%%%%%%%%%%%%%%%%%%%%%%%%%%%%%%
\subsection{Path-width}\label{sec-pw}
In 1983, three years ahead of their definition for tree-width,
Robertson and Seymour introduced the notion of path-width~\cite{RS83}.

\begin{definition}[Path-width]
  A {\em path-decomposition} of a graph $G$ is a sequence $(X_1,
  \ldots, X_r)$ of subsets of $V(G)$, called {\em bags}, such that the
  subsequent three conditions hold:

  \begin{description}
  \item[{\textnormal{(pw-1)}}]
    $\bigcup_{1 \leq i \leq r} X_i = V(G)$,\footnote{Please note that
      (pw-1) has no influence on the width of a path-decomposition, see
      footnote \ref{fn1}.}
  \item[{\textnormal{(pw-2)}}]
    for every edge $\{u,v\} \in E(G)$ there is a set $X_i$, $1 \leq i
    \leq r$, with $u, v \in X_i$, and
  \item[{\textnormal{(pw-3)}}]
    for all $i, j, \ell$ with $1 \leq i < j < \ell \leq r$ it holds
    that $X_i \cap X_\ell \subseteq X_j$.
  \end{description}

  The {\em width} of a path-decomposition $(X_1, \ldots, X_r)$ is
  defined as
  \begin{align*}
    \max_{1 \leq i \leq r} |X_i|-1.
  \end{align*}
  Finally, the {\em path-width} of a graph $G$, denoted by $\pws(G)$,
  is the smallest integer $k$, such that there is a path-decomposition
  $(X_1, \ldots, X_r)$ for $G$ of width $k$.
\end{definition}

Similar to the tree-width of a graph, the path-width of a graph $G$ is
limited by $|V(G)| - 1$, as all vertices of $G$ can be stored in a
single bag.
However, deciding whether the path-width of a graph is at most some
given integer $w$ is NP-complete~\cite{KF79}.
Even for special graph classes, such as bipartite graphs, complements
of bipartite graphs~\cite{ACP87}, chordal graphs~\cite{Gus93},
bipartite distance hereditary graphs~\cite{KBMK93}, and planar graphs
with maximum vertex degree 3~\cite{MS88}, it is NP-complete to decide
if the path-width is at most a given integer $w$.
On the other hand, there are also special graph classes, such as
permutation graphs~\cite{BKK93}, circular arc graphs~\cite{ST07a}, and
co-graphs~\cite{BM93}, for which one can decide in polynomial time
whether the path-width is at most a given integer $w$.

\begin{example}[Path-decomposition]\label{ex-pw}
  $\calX = (\{ a, b, c \}, \{ c, f, g \}, \{ c, d, e \})$ is a
  path-de\-com\-po\-si\-ti\-on
  of width $2$ for the graph shown in Figure~\ref{F-ex-tw}.
\end{example}

Besides the definition of path-width by Robertson and Seymour, there
are further definitions of path-width, for example by vertex
separation number~\cite{Kin92}.
Again, we refer to the work by Bodlaender~\cite{Bod98} for an overview
of equivalent definitions.

With a small adaption to the proof for Lemma~\ref{le-tw-deg-v}, we
obtain a similar result for path-width.

\begin{lemma}\label{le-pw-deg-v}
  Every graph of path-width at most $k$ has a vertex of degree at most
  $k$.
\end{lemma}

Having this lemma, we obtain the subsequent proposition for path-width
with the same proof as used for Proposition~\ref{prop-tw-edges},
adapted to path-width.

\begin{proposition}
  Let $G$ be a graph.
  Then, it holds that
  \begin{align*}
    |E(G)| \leq \pws(G) |V(G)| - \frac{1}{2} \pws(G) (\pws(G) + 1).
  \end{align*}
\end{proposition}

The same work by Kneis et al.~\cite{KMRR09} that provides an upper
bound for the tree-width of a graph in terms of its edges and vertices
also provides an upper bound for the path-width of a graph in terms of
its edges and vertices.

\begin{proposition}
  Let $G$ be a graph.
  Then, it holds that
  \begin{align*}
    \pws(G)\leq \frac{|E(G)|}{5.769} + \bigo(\log(|V(G)|)).
  \end{align*}
\end{proposition}

Following the same argumentation as for Lemma~\ref{le-tw-deg-v}, we obtain
the subsequent result for the path-width of $k$-connected graphs.

\begin{lemma}\label{le-pw-con}
  Every $k$-connected graph has path-width at least $k$.
\end{lemma}

As every path is a tree, every path-decomposition $\calX = (X_1,
\ldots, X_r)$ can be interpreted as a tree-decomposition, such that
the results from Lemma~\ref{le-tw-clique} and~\ref{le-tw-bip-clique}
also hold for path-decompositions.
Subsequently, we provide reformulations of the results by Bodlaender
and M\"ohring~\cite{BM93}, adjusted for path-width.

\begin{lemma}[Clique containment]\label{le-pw-clique}
  Let $\calX = (X_1, \ldots, X_r)$ be a path-decomposition for a graph
  $G$ and $C \subseteq V(G)$ a clique in $G$.
  Then, there exists some $1 \leq i \leq r$ with $C \subseteq X_i$.
\end{lemma}

\begin{lemma}[Complete bipartite subgraph
  containment]\label{le-pw-bip-clique}
  Let $\calX = (X_1, \ldots, X_r)$ be a path-decomposition for a graph $G$
  and $A, B \subseteq V(G)$ with $A \cap B = \emptyset$ and
  $\{ \{ u, v \} \mid u \in A, v \in B \} \subseteq E(G)$.
  Then, there exists some $1 \leq i \leq r$ with $A \subseteq X_i$ or
  $B \subseteq X_i$.
\end{lemma}

%%%%%%%%%%%%%%%%%%%%%%%%%%%%%%%%%%%%%%%%%%%%%%%%%%%%%%%%%%%%%%%%%%%%%%%%%%%
\subsection{Graph Parameters and Relations}
In this subsection, we provide results setting the parameters
tree-width and path-width into relation with each other as well as
with other, well-known graph parameters, such as
\begin{enumerate}
\item the {\em independence number} $\alpha(G)$, specifying the size of a
  largest independent set in $G$,
\item the {\em clique number} $\omega(G)$, specifying the size of a largest
  clique in $G$, and
\item the {\em chromatic number} $\chi(G)$, specifying the least number of
  distinct colors required to color all vertices of $G$, such that no
  two adjacent vertices obtain the same color.
\end{enumerate}

With the subsequent theorem, we prove that the tree-width of a graph
is always bounded by the path-width of the graph, whereas the
path-width of a graph cannot be bounded only in the graph's
tree-width, but in the tree-width and the number of vertices.

\begin{theorem}\label{pw-tw}
  For a graph $G$ it holds that
    \begin{align*}
      \tws(G) \leq \pws(G) \quad \text{and} \quad \pws(G) \leq
      (\tws(G) + 1) (\log_3(2|V(G)| + 1) + 1) - 1.
   \end{align*}
\end{theorem}

\begin{proof}
  The first relation follows by the definitions of tree-width and
  path-width, as every path-decomposition can also be interpreted as a
  tree-decomposition.

  The second relation, making use of the idea of Theorem~6 in the work
  by Korach and Solel~\cite{KS93}, can be shown as follows.
  Let $(\calX, T)$ be a tree-decomposition for $G$ and denote its bags
  by $X_u \in \calX$ for $u \in V(T)$.
  According to the result (10.16) in the book by Kleinberg and
  Tardos~\cite{KT04}, we can assume $(\calX, T)$ to be a non-redundant
  tree-decomposition for $G$, i.e., $|V(T)| \leq |V(G)|$ holds.
  Following Scheffler~\cite{Sch89}, the path-width of any tree $S$ can be limited from above by
  $\pws(S) \leq \log_3(2 |V(S)| + 1)$.
  Thus, since $(\calX, T)$ is a non-redundant tree-decomposition for
  $G$, it follows that there exists a path-decomposition $\mathcal{Y}$
  for $T$ of path-width $\pws(T) \leq \log_3(2 |V(G)| + 1)$.
  We denote the bags of $\calY$ by $Y_i \in \calY$ for $1 \leq i \leq
  r$.
  Given this path-decomposition $\calY$ for $T$, we construct $\calZ =
  (Z_1, \ldots, Z_r)$ with $Z_i = \bigcup_{u \in Y_i} X_u$.
  By Theorem~6 in the work by Korach and Solel~\cite{KS93}, 
  it follows that $\calZ$ is a path-decomposition for $G$.  We
  know that $|X_u| \leq \tws(G) + 1$ holds for all $u \in V(T)$ and
  $|Y_i| \leq \log_3(2 |V(G)| + 1) + 1$ for $1 \leq i \leq r$.
  Consequently, it follows that $|Z_i| \leq (\tws(G) + 1)(\log_3(2
  |V(G)| + 1) + 1)$ is true for $1 \leq i \leq r$, such that we obtain
  $\pws(G) \leq (\tws(G) + 1)(\log_3(2 |V(G)| + 1) + 1) - 1$.
\end{proof}

By Lemma~\ref{le-tw-clique} and Lemma~\ref{le-pw-clique} we know that
for a complete graph with $n \in \N$ vertices, $K_n$, it holds that
$\tws(K_n) = n - 1 = \pws(K_n)$.
Consequently, the first inequality stated in Theorem~\ref{pw-tw} is
tight.
The second inequality is asymptotically optimal, as trees have
tree-width one while their path-width grows asymptotically in their
height, which is at least logarithmic in the number of vertices.

The same two lemmas imply that for a graph $G$ every clique of $G$ is
subset of at least one bag of every tree-decomposition and
path-decomposition for $G$.
With this fact at hand, the subsequent proposition follows immediately.

\begin{proposition}
  For a graph $G$ it holds that
    \begin{align*}
  \omega(G)-1 \leq \tws(G)  \quad \text{and} \quad \omega(G)-1 \leq \pws(G).
   \end{align*}
\end{proposition}

Chleb\'{i}kov\'{a}~\cite{CHL02} proved the subsequent proposition's first bound, i.e., the one
for tree-width.
To do so, he argued that by splitting a graph $G$'s vertices into an
independent set $I$ of size $\alpha(G)$ and $|V(G)| - \alpha(G)$
subsets $V_i \subseteq V(G) \setminus I$ of size one, it follows that
$G$ is a subgraph of a complete $(|V(G)|-\alpha(G)+1)$-partite graph.
Then, it is easy to see that the bags $X_v = (V(G) \setminus I) \cup
\{ v \}$ for $v \in I$ allow a tree-decomposition of width $|V(G)|
- \alpha(G)$ for the $(|V(G)|-\alpha(G)+1)$-partite graph.
As the same bags $X_v$, $v \in I$, allow a path-decomposition as
well, the same bound follows for path-width.

\begin{proposition}
  For a graph $G$ it holds that
      \begin{align*}
        \alpha(G) + \tws(G) \leq |V(G)| \quad \text{and} \quad
        \alpha(G) + \pws(G) \leq |V(G)|.
     \end{align*}
\end{proposition}

Finally, Chleb\'{i}kov\'{a}~\cite{CHL02} also proved the subsequent proposition's first bound.
Making use of the first bound in Theorem~\ref{pw-tw}, we extended this
result to path-width, too.

\begin{proposition}\label{prop-chi}
  For a graph $G$ it holds that
    \begin{align*}
      \chi(G) \leq \tws(G) + 1\quad \text{and} \quad \chi(G) \leq
      \pws(G) + 1.
     \end{align*}
\end{proposition}

%%%%%%%%%%%%%%%%%%%%%%%%%%%%%%%%%%%%%%%%%%%%%%%%%%%%%%%%%%%%%%%%%%%%%%%%%%%
%%%%%%%%%%%%%%%%%%%%%%%%%%%%%%%%%%%%%%%%%%%%%%%%%%%%%%%%%%%%%%%%%%%%%%%%%%%
%%%%%%%%%%%%%%%%%%%%%%%%%%%%%%%%%%%%%%%%%%%%%%%%%%%%%%%%%%%%%%%%%%%%%%%%%%%
\section{Unary Graph Operations and Graph Transformations}\label{sec-un}

Let $G$ be a non-empty graph and $f$ a unary graph transformation that
creates a new graph $f(G)$ from $G$.
In this section of our work, we consider the tree-width and path-width
of graph $f(G)$ with respect to the tree-width and path-width of $G$.
Especially, we consider the graph transformations
vertex deletion,
vertex addition,
edge deletion
edge addition,
taking a subgraph,
edge subdivision,
vertex identification,
edge contraction,
taking a minor,
powers of graphs,
line graphs,
edge complements,
local complements,
Seidel switching,
and Seidel complementation.

%%%%%%%%%%%%%%%%%%%%%%%%%%%%%%%%%%%%%%%%%%%%%%%%%%%%%%%%%%%%%%%%%%%%%%%%%%%
\subsection{Vertex Deletion and Vertex Addition\label{vertexadd}}

\paragraph{Vertex Deletion}
Let $G$ be a graph and $v \in V(G)$ a vertex of $G$.
By $G - v$ we denote the graph we obtain from $G$ when removing
$v$ and all its incident edges, i.e.,
\begin{align*}
  G - v = (V(G) \setminus \{v\}, E(G) \setminus \{ \{v, u\} \mid u \in
  N_G(v)\}).
\end{align*}

Having the graph operation of {\it vertex deletion} defined, subsequently
we consider the tree-width and path-width of $G - v$.

\begin{theorem}\label{del-v}
  \label{Tvertexrem}
  For a graph $G$ and vertex $v \in V(G)$, it holds that
  \begin{align*}
    \tws(G) - 1 & \le \tws(G - v) \le \tws(G) \quad \text{and} \\
    \pws(G) - 1 & \le \pws(G - v) \le \pws(G).
  \end{align*}
\end{theorem}

\begin{proof}
  Removing $v$ from every bag of a tree-decomposition
  (path-decomposition) for $G$ and deleting all resulting, empty bags, we
  obtain a tree-decomposition (path-decomposition) for $G - v$.
  Consequently, we obtain $\tws(G - v) \leq \tws(G)$ and $\pws(G - v)
  \leq \pws(G)$.
  Adding $v$ and its incident edges to $G - v$ results in $G$, such
  that the lower bounds follow from the upper bounds of Theorem~\ref{Tvertex}.
\end{proof}

%%%%%%%%%%%%%%%%%%%%%%%%%%%%%%%%%%%%%%%%%%%%%%%%%%%%%%%%%%%%%%%%%%%%%%%%%%%
\paragraph{Vertex Addition}
Let $G$ be a graph, $N \subseteq V(G)$ a subset of vertices from $G$,
and $v \not\in V(G)$ a newly introduced vertex.
By $G +_N v$ we denote the graph we obtain from $G$ when inserting
$v$ with neighborhood $N_G(v) = N$, i.e.,
\begin{align*}
  G +_N v = (V(G) \cup \{v\}, E(G) \cup \{ \{v, u\} \mid u \in N \}).
\end{align*}

In the special case, when $N_G(v) = \{ u \}$ holds for a vertex $u \in
V(G)$, we call $v$ a {\em pendant vertex} of $G$.
If $N_G(v) = V(G) \setminus \{ v \}$ is true, we call $v$ a {\em
dominating vertex} of $G$.
With {\it vertex addition} defined, we consider the tree-width and
path-width of graph $G +_N v$ in the ensuing theorem.

\begin{theorem}\label{Tvertex}
  For a graph $G$, a subset of vertices $N \subseteq V(G)$, and a
  vertex $v \not\in V(G)$ it holds that
  \begin{align*}
    \tws(G) & \le \tws(G +_N v) \le \tws(G) + 1 \quad \text{and} \\
    \pws(G) & \le \pws(G +_N v) \le \pws(G) + 1.
  \end{align*}
\end{theorem}

\begin{proof}
  Introducing $v$ to all bags of a tree-decomposition
  (path-decomposition) of $G$, we obtain a tree-decomposition
  (path-decomposition) of $G +_N v$.
  Thereby, it follows that $\tws(G +_N v) \leq \tws(G) + 1$ and $\pws(G
  +_N v) \leq \pws(G) + 1$ is true.
  Removing $v$ from $G +_N v$, we obtain $G$.
  Consequently, the lower bounds follow from the upper bounds of
Theorem~\ref{del-v}.
\end{proof}

We can always add a pendant vertex to a graph without increasing the
graph's tree-width.
To do so, we introduce a new bag that contains the new vertex and its
solely neighbor.
Afterwards, the new bag is connected to any bag of a tree-decomposition
for the graph, which already contains the neighbor.

\begin{corollary}\label{cor-pv}
  For a graph $G$, a vertex $u \in V(G)$, and a newly introduced
  vertex $v \not\in V(G)$, it holds that
  \begin{align*}
    \tws(G +_{\{ u \}} v) = \max(\tws(G), 1).
  \end{align*}
\end{corollary}

Contrarily, the subsequent example shows that the previous statement
does not hold with respect to path-width.
Introducing a pendant vertex to a graph can increase its path-width.

\begin{example}
  The path-decomposition $\calX = (\{a, b\}, \{b, c\}, \{c, f\}, \{c,
  d\}, \{d, e\})$ shows that the path-width of graph $C$ in
  Figure~\ref{F-co} is one.
  Introducing pendant vertex $g$ to $C$, as depicted in graph
  $I(K_{1,3})$ of the same figure, the path-width increases to two,
  $\calX' = (\{ a, b\}, \{ b, c\}, \{c, f, g\}, \{ c, d\}, \{d, e\})$.
\end{example}

%%%%%%%%%%%%%%%%%%%%%%%%%%%%%%%%%%%%%%%%%%%%%%%%%%%%%%%%%%%%%%%%%%%%%%%%%%%
\subsection{Edge Addition and Edge Deletion}\label{edgeadd}

\paragraph{Edge Deletion}
Let $G$ be a graph and $v, w \in V(G)$ two vertices.
For $\{v, w\} \in E(G)$, we define by $G - \{v,w\}$ the graph we obtain
from $G$ by deleting edge $\{v,w\}$, i.e.,
\begin{align*}
  G - \{v, w\} = (V(G), E(G) \setminus \{\{v, w\}\}).
\end{align*}

With {\it edge deletion} defined, the subsequent theorem shows that
removing an edge from a graph decreases the width of the graph at most
by one.

\begin{theorem}
  \label{Tedge-rem}
  For a graph $G$ and two different vertices $v, w \in V(G)$ it holds
  that
  \begin{align*}
    \tws(G) - 1 & \le \tws(G - \{v, w\}) \le \tws(G) \quad \text{and} \\
    \pws(G) - 1 & \le \pws(G - \{v, w\}) \le \pws(G).
  \end{align*}
\end{theorem}

\begin{proof}
  The upper bound follows immediately, because a tree-decomposition
  (path-decomposition) for $G$ is also a tree-decomposition
  (path-decomposition) for $G - \{v,w\}$.

  As $G$ can be obtained from $G - \{v,w\}$ by adding edge $\{v,w\}$,
  the lower bound follows from the upper bound of Theorem~\ref{Tedge}.
\end{proof}

%%%%%%%%%%%%%%%%%%%%%%%%%%%%%%%%%%%%%%%%%%%%%%%%%%%%%%%%%%%%%%%%%%%%%%%%%%%
\paragraph{Edge Addition}
Let $G$ be a graph and $v, w \in V(G)$ two vertices.
For $\{v, w\} \not\in E(G)$, we define by $G + \{v, w\}$ the graph we obtain
from $G$ when adding edge $\{v, w\}$, i.e.,
\begin{align*}
  G + \{v, w\} = (V(G), E(G) \cup \{ \{v, w\} \}).
\end{align*}

Having {\it edge addition} defined, our next theorem shows that
inserting an edge into a graph increases the graph's width at most by
one.

\begin{theorem}
  \label{Tedge}
  For a graph $G$ and two different vertices $v,w \in V(G)$ it holds that
  \begin{align*}
    \tws(G) & \le \tws(G + \{v, w\}) \le \tws(G) + 1 \quad \text{and} \\
    \pws(G) & \le \pws(G + \{v, w\}) \le \pws(G) + 1.
  \end{align*}
\end{theorem}

\begin{proof}
  Given a tree-decomposition (path-decomposition) for $G$, we obtain a
  tree-decomposition (path-decomposition) for $G + \{v, w\}$ by adding
  one of the two vertices, $v$ or $w$, to all its bags.
  Consequently, $\tws(G + \{v,w\}) \leq \tws(G) + 1$ and $\pws(G +
  \{v,w\}) \leq \pws(G) + 1$ hold.

  The lower bounds follow by the fact that a tree-decomposition
  (path-decomposition) for $G + \{v, w\}$ is also a tree-decomposition
  (path-decomposition) for $G$.
\end{proof}

%%%%%%%%%%%%%%%%%%%%%%%%%%%%%%%%%%%%%%%%%%%%%%%%%%%%%%%%%%%%%%%%%%%%%%%%%%%
\subsection{Subgraph}\label{sec-sg}
So far we have only studied unary graph operations within this section of
our work.
However, in this subsection, we deviate from this path and study the
unary graph transformation of {\it taking a subgraph}.
This act of modifying a graph is not deterministic, since there are
various subgraphs one can take from any given graph and it is not
explicitly defined which one should be taken.
Consequently, taking a subgraph of a graph multiple times can result
in different subgraphs, making the studied modification a graph
transformation but no graph operation.

Note, that taking a subgraph of a graph can be interpreted as a
sequence of vertex deletion and edge deletion operations.
Hence, every subgraph of a graph can be obtained by deleting selected
vertices and edges from the original graph, such that the subsequent
corollary follows immediately by Theorem~\ref{del-v} and
Theorem~\ref{Tedge-rem}.

\begin{corollary}\label{cor-pw-tw-subgr}
  For a graph $G$ and any subgraph $H$ of $G$ it holds that $\tws(H)
  \leq \tws(G)$ and $\pws(H) \leq \pws(G)$.
\end{corollary}

%%%%%%%%%%%%%%%%%%%%%%%%%%%%%%%%%%%%%%%%%%%%%%%%%%%%%%%%%%%%%%%%%%%%%%%%%%%
\subsection{Vertex Identification}
For a graph $G$ and two different vertices $v,w \in V(G)$, the
{\em identification} of $v$ and $w$ in $G$, denoted by $\ident(G, v,
w)$, consists of vertex set $(V(G) \setminus \{v,w\}) \cup \{u\}$ and
edge set
\begin{align*}
  E(G) & \setminus \{\{x, y\} \mid x \in V(G), y \in \{v, w\}\} \\
       & \cup \{\{x, u\} \mid x \in (N_G(v) \cup N_G(w)) \setminus \{v,w\} \}.
\end{align*}
Given this definition of vertex identification, the ensuing result
summarizes the graph operation's effect on the tree-width and
path-width of the involved graph.

\begin{theorem}\label{Tid}
  For a graph $G$ and two different vertices $v, w \in V(G)$ it holds that
  \begin{align*}
    \tws(G) - 1 & \le \tws(\ident(G, v, w)) \le \tws(G) + 1 \quad \text{and} \\
    \pws(G) - 1 & \le \pws(\ident(G, v, w)) \le \pws(G) + 1.
  \end{align*}
\end{theorem}

\begin{proof}
  For the upper bounds, let $(\calX, T)$ be a tree-decomposition
  ($\calX$ be a path-decomposition) for $G$ of width $\tws(G)$
  ($\pws(G)$).
  To obtain a tree-decomposition (path-decomposition) for
  $\ident(G,v,w)$, we proceed as follows.
  In a first step, replace all occurrences of $v$ and $w$ in all bags
  of $\calX$ by $u$ and denote the result by $\calX'$.
  Since $\{v,w\}$ is not necessarily an edge of $G$, $\calX'$ could
  violate (tw-3) ((pw-3)), i.e., the bags of $\calX'$ containing $u$
  might not be connected.
  In this case, we add $u$ to all bags between the disconnected
  components.
  With that, $(\calX', T)$ is a valid tree-decomposition ($\calX'$ is
  a valid path-decomposition) for $\ident(G,v,w)$.
  To obtain $\calX'$ from $\calX$, we increased the width at most by
  one, such that $\tws(\ident(G, v, w)) \leq \tws(G) + 1$
  ($\pws(\ident(G, v, w)) \leq \pws(G) + 1$) follows.

  For the lower bounds, we proceed as follows. First, rename vertex
  $u$ of $\ident(G,v,w)$ to $v$ and denote the resulting graph by $G'$.
  Next, add a new vertex $w$ with neighborhood $N_G(w)$ to $G'$.
  By Theorem~\ref{Tvertex} it follows that $\tws(G' +_{N_G(w)} w) \leq
  \tws(G') + 1$ and $\pws(G' +_{N_G(w)} w) \leq \pws(G') + 1$ are true.
  Since $G$ is a subgraph of $G' +_{N_G(w)} w$, with
  Corollary~\ref{cor-pw-tw-subgr} we obtain that
  \begin{align*}
    \tws(G) & \leq \tws(G' +_{N_G(w)} w) \leq
    \tws(\ident(G,v,w)) + 1 \quad \text{and} \\
    \pws(G) & \leq \pws(G' +_{N_G(w)} w) \leq
    \pws(\ident(G,v,w)) + 1
  \end{align*}
  hold, which yields the lower bounds $\tws(G) - 1 \leq
  \tws(\ident(G,v,w))$ and $\pws(G) - 1 \leq \pws(\ident(G,v,w))$.
\end{proof}

Note that the upper bounds specified in the previous theorem
are tight, since a vertex identification on the end vertices of a path
results in a cycle with increased tree-width and path-width.

%%%%%%%%%%%%%%%%%%%%%%%%%%%%%%%%%%%%%%%%%%%%%%%%%%%%%%%%%%%%%%%%%%%%%%%%%%%
\subsection{Edge Contraction}\label{contr}

In the case that the two vertices $v, w \in V(G)$ of a vertex identification
$\ident(G,v,w)$ are adjacent, i.e., $\{v, w\} \in E(G)$, we call the
operation an {\em edge contraction}.

\begin{theorem}\label{Tid-ad}
  For a graph $G$ and two different vertices $v, w \in V(G)$ with
$\{v, w\} \in E(G)$ it holds that
  \begin{align*}
    \tws(G) - 1 & \le \tws(\ident(G, v, w)) \le \tws(G) \quad \text{and} \\
    \pws(G) - 1 & \le \pws(\ident(G, v, w)) \le \pws(G).
  \end{align*}
\end{theorem}

\begin{proof}
  Let $(\calX, T)$ be a tree-decomposition ($\calX$ be a
  path-decomposition) for $G$ of width $\tws(G)$ ($\pws(G)$).
  We replace all occurrences of $v$ and $w$ in all bags of
  $\calX$ by $u$ and denote the resulting decomposition by
  $\calX'$.
  Since $v$ and $w$ are adjacent in $G$, we know
  by (tw-2) ((pw-2)) that there is at least one bag in $\calX$
  which contains $v$ and $w$.
  Consequently, (tw-3) ((pw-3)) must hold for $\calX'$ and it follows
  that $(\calX', T)$ is a valid tree-decomposition ($\calX'$ is a valid
  path-decomposition) for $\ident(G,v,w)$ of width at most $\tws(G)$
  ($\pws(G)$).\footnote{
    Because $\{ v, w \} \in E(G)$ is true, we can argue that (tw-3)
    ((pw-3)) holds for $\calX'$.
    This argument is not valid for an arbitrary vertex identification,
    as $v$ and $w$ are not guaranteed to be adjacent, see
    Theorem~\ref{Tid}.}

  The lower bounds follow by the same argumentation as for the lower
  bounds in the proof of Theorem~\ref{Tid}.
\end{proof}

Contracting any edge of a clique $K_n$ of size $n$ results in a clique
$K_{n-1}$ of size $n - 1$.
From Lemma~\ref{le-tw-clique} (Lemma~\ref{le-pw-clique}) we know that
$\tws(K_n) = n - 1$ ($\pws(K_n) = n - 1$) holds.
Consequently, the lower bounds specified in the previous theorem are
tight.

%%%%%%%%%%%%%%%%%%%%%%%%%%%%%%%%%%%%%%%%%%%%%%%%%%%%%%%%%%%%%%%%%%%%%%%%%%%
\subsection{Edge Subdivision}

Let $G$ be a graph, $u \not\in V(G)$ a newly introduced vertex, and
$\{v, w\} \in E(G)$ an edge of $G$.
The {\em edge subdivision} of $\{v, w\}$ in $G$, denoted by $\subdiv(G, v,
w)$, consists of vertex set $V(G) \cup \{u\}$ and edge set $(E(G)
\setminus \{\{v, w\}\}) \cup \{\{v, u\}, \{u, w\}\}$.

With this definition at hand, the subsequent theorem states the effect
of an edge subdivision on the tree-width and path-width of a given
graph.

\begin{theorem}
  \label{Tsub}
  For a graph $G$ and an edge $\{v, w\} \in E(G)$ it holds that
  \begin{align*}
    \tws(\subdiv(G,v,w)) & = \tws(G) \quad \text{and} \\
    \pws(G) \le \pws(\subdiv(G,v,w)) & \le \pws(G) + 1.
  \end{align*}
\end{theorem}

\begin{proof}
  $G$ is isomorphic to $\ident(\subdiv(G, v, w), u, w)$, such that the
  upper bounds of Theorem~\ref{Tid-ad} yield
  \begin{align*}
    \tws(G) & = \tws(\ident(\subdiv(G, v, w), u, w)) \leq
              \tws(\subdiv(G, v, w)) \quad \text{and} \\
    \pws(G) & = \pws(\ident(\subdiv(G, v, w), u, w)) \leq
              \pws(\subdiv(G, v, w)),
  \end{align*}
  resulting in this theorem's lower bounds.

  For the upper bound of tree-width, let us distinguish the subsequent
  two cases.
  \begin{description}
  \item [Case~1:] $\tws(G) = 1$

    In this case, $G$ is a forest.
    Since an edge subdivision does not alter
    this fact, $\subdiv(G,v,w)$ is still a forest and
    $\tws(\subdiv(G,v,w)) = 1$ must hold.
  \item [Case~2:] $\tws(G) > 1$

    In this case, the biggest bag of every
    tree-decomposition of $G$ contains at least three vertices.
    Furthermore, by (tw-2) it follows that in every tree-decomposition
    of $G$ there is at least one bag which contains $v$ and $w$.
    Let us denote this bag by $X$.
    Adding a new bag $X'$ with vertices $\{ u, v, w \}$ and making
    it adjacent to $X$ results in a tree-decomposition of
    $\subdiv(G,v,w)$ with unaltered size, such that
    $\tws(\subdiv(G,v,w)) \leq \tws(G)$ follows.
  \end{description}

  For the upper bound of path-width, we proceed as follows.
  By (pw-2), we know that in every path-decomposition of $G$ there
  exists a bag containing $v$ and $w$.
  Adding $u$ to this bag, we obtain a valid path-decomposition for
  $\subdiv(G, v, w)$.
  Consequently, $\pws(\subdiv(G,v,w)) \leq \pws(G) + 1$ follows.
\end{proof}

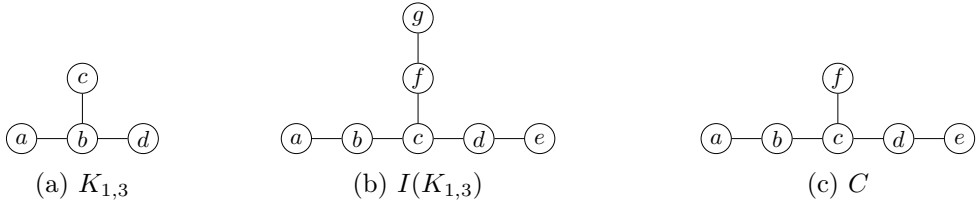
\begin{figure}
  \centering
  \begin{subfigure}{0.15\textwidth}
    \centering
    \scalebox{0.8}{
      \begin{tikzpicture}
        \draw (-1,0) \gnode (a) {$a$};
        \draw (0,0) \gnode (b) {$b$};
        \draw (0,1) \gnode (c) {$c$};
        \draw (1,0) \gnode (d) {$d$};
        \draw[-] (a) to (b);
        \draw[-] (b) to (c);
        \draw[-] (b) to (d);
      \end{tikzpicture}
    }
    \caption{$K_{1,3}$}
  \end{subfigure}
  \hspace{5ex}
  \begin{subfigure}{0.3\textwidth}
    \centering
    \scalebox{0.8}{
      \begin{tikzpicture}
        \draw (-2,0) \gnode (a) {$a$};
        \draw (-1,0) \gnode (b) {$b$};
        \draw (0,0) \gnode (c) {$c$};
        \draw (1,0) \gnode (d) {$d$};
        \draw (2,0) \gnode (e) {$e$};
        \draw (0,1) \gnode (f) {$f$};
        \draw (0,2) \gnode (g) {$g$};
        \draw[-] (a) to (b);
        \draw[-] (b) to (c);
        \draw[-] (c) to (f);
        \draw[-] (f) to (g);
        \draw[-] (c) to (d);
        \draw[-] (d) to (e);
      \end{tikzpicture}
    }
    \caption{$I(K_{1,3})$}
  \end{subfigure}
  \hspace{5ex}
  \begin{subfigure}{0.3\textwidth}
    \centering
    \scalebox{0.8}{
      \begin{tikzpicture}
        \draw (-2,0) \gnode (a) {$a$};
        \draw (-1,0) \gnode (b) {$b$};
        \draw (0,0) \gnode (c) {$c$};
        \draw (1,0) \gnode (d) {$d$};
        \draw (2,0) \gnode (e) {$e$};
        \draw (0,1) \gnode (f) {$f$};
        \draw[-] (a) to (b);
        \draw[-] (b) to (c);
        \draw[-] (c) to (f);
        \draw[-] (c) to (d);
        \draw[-] (d) to (e);
      \end{tikzpicture}
    }
    \caption{$C$}
  \end{subfigure}
  \caption{Special graphs.}
  \label{F-co}
\end{figure}

The upper bound for path-width given in the previous theorem is tight.
The path-width of the caterpillar\footnote{A {\em caterpillar} graph
  is a tree for which the removal of all pendant vertices results in a
  chordless path.} graph $C$ in Figure~\ref{F-co} equals one.
Contrarily, the path-width of the graph obtained by subdividing edge
$\{c,f\}$ of $C$, depicted as $I(K_{1,3})$ in Figure~\ref{F-co}, is two.

Latest after subdividing all edges of a graph, the resulting graph
must be bipartite.
The resulting graph, one obtains after subdividing all edges of a
graph $G$, is called the {\em incidence graph} of $G$, denoted by
$I(G)$.
The subsequent corollary provides bounds for a graph's incident graph
and follows from Theorem~\ref{Tsub}.

\begin{corollary}
  \label{cor-inci}
  For a graph $G$ it holds that
  \begin{align*}
    \tws(I(G)) & = \tws(G) \quad \text{and} \\
    \pws(G) \leq \pws(I(G)) & \leq \pws(G) + 1.
  \end{align*}
\end{corollary}

\begin{proof}
  The result for tree-width follows directly by Theorem~\ref{Tsub}.

  Contracting all subdivided edges of $I(G)$, we obtain $G$. By
  Theorem~\ref{Tid-ad} we know that edge contractions do no increase
  the resulting graph's path-width, such that $\pws(G) \leq
  \pws(I(G))$ follows.

  Denote by $\calX$ a path-decomposition for $G$ of width $\pws(G)$ and
  let $\{v,w\} \in E(G)$ be an edge of $G$ as well as $x_{\{v,w\}}$ the
  node introduced to subdivide the edge.
  By (pw-2) we know that there is a bag $X$ in $\calX$ with $v,w
  \in X$.
  We introduce a new bag $X' = X \cup \{ x_{\{ v, w\}} \}$ to $\calX$
  as successor of $X$ and denote the resulting decomposition $\calX'$.
  Then, $\calX'$ is a path-decomposition for $\subdiv(G, v, w)$ of
  width $\pws(G) + 1$.
  We can repeat this step for every edge of $G$ and finally, obtain a
  path-decomposition for $I(G)$ of width $\pws(G) + 1$, such that
  $\pws(I(G)) \leq \pws(G) + 1$ follows.
\end{proof}

%%%%%%%%%%%%%%%%%%%%%%%%%%%%%%%%%%%%%%%%%%%%%%%%%%%%%%%%%%%%%%%%%%%%%%%%%%%
\subsection{Minor}\label{subsection-minor}
Every graph $H$ one may obtain from a given graph $G$, by applying a
finite sequence of arbitrary edge deletion and edge contraction
operations, is called a {\em minor} of $G$.
Obviously, taking a minor is a graph transformation but no graph
operation, as the applied sequence of operations is not deterministic.

Given this definition, the subsequent theorem follows immediately from
Theorem~\ref{Tedge-rem} and Theorem~\ref{Tid-ad}.

\begin{theorem} \label{mi-tw-pw}
  For a graph $G$ and a minor $H$ of $G$, it holds that
  $\tws(H) \leq \tws(G)$ and $\pws(H) \leq \pws(G)$.
\end{theorem}

In other words, the fact that a graph has tree-width (path-width) at
most $k \in \N$ is preserved under the graph transformation of taking
a minor.

Subsequently, we cite one of the probably most important theorems in
graph theory, the {\em minor theorem}, shown by Robertson and
Seymour~\cite{RS04}.

\begin{theorem}[Minor theorem]\label{cor-minor-satz-x}
  In every infinite set of graphs there are two graphs, such that one
  of them is a minor of the other.
\end{theorem}

Before the minor theorem was proven by Robertson and Seymour, it was
known as Wagner's conjecture~\cite{W70}.
The following theorem is an important corollary of the minor
theorem~\cite{RS04}.

\begin{theorem}\label{cor-minor-satz}
  Every set of graphs that is closed under minors can be defined by a
  finite set of forbidden minors.
\end{theorem}

For $k \in \N$, Theorem~\ref{mi-tw-pw} implies that the set of graphs
with tree-width (path-width) at most $k$ is closed under
minors.
Consequently, Theorem~\ref{cor-minor-satz} then implies that
the set of graphs with tree-width (path-width) at most $k$ can be
characterized by a finite set of forbidden minors.
For small values of $k$ these sets of forbidden minors are known.
For example, Kinnersley and Langston~\cite{KL94} characterized the set
of all graphs with path-width at most one by the forbidden minors
$K_3$ and $I(K_{1,3})$, c.f. Figure~\ref{F-co}.
Furthermore, they characterized the set of graphs with
path-width at most two by $110$ forbidden minors~\cite{KL94}.
The set of all graphs with tree-width at most one is characterized by
the forbidden minor $K_3$ and the set of all graphs with tree-width at
most two is characterized by the forbidden minor $K_4$, see the work
by Bodlaender and van Antwerpen-de Fluiter~\cite{BV01}.
For the four forbidden minors that characterize the set of all graphs
with tree-width at most three, see the work by Arnborg, Proskurowski,
and Corneil~\cite{APC90}.

The main algorithmic consequence of the minor theorem is stated in the
subsequent theorem, the {\em minor test}, also shown by Robertson and
Seymour~\cite{RS95}.

\begin{theorem}[Minor test]\label{text-m}
  For a fixed graph $H$ and a given graph $G$, one can decide in time
  in $\bigo(|V(G)|^3)$ whether $H$ is a minor of $G$.\footnote{A more
    precise formulation of the runtime would be $\bigo(f(k)|V(G)|^3)$ for
    some function $f$ and $k \leq \text{size}(H)$.}
\end{theorem}

From a parameterized point of view, the minor test is fixed-parameter
tractable with respect to the parameter $\text{size}(H)$.
However, the constant behind the $\bigo$-notation in
Theorem~\ref{text-m} depends on the parameter $\text{size}(H)$ and is
huge.
Kawarabayashi, Kobayashi, and Reed~\cite{KKR12} showed an improved
version of the previous result, allowing to decide the minor test in time
in $\bigo(|V(G)|^2)$.

The ensuing corollary follows immediately by
Theorem~\ref{cor-minor-satz} and the application of Theorem~\ref{text-m}
for every forbidden minor.

\begin{corollary}\label{cor-recog}
  For every set of graphs $\calG$ that is closed under minors and
  every graph $G$ one can decide in time polynomial in the size of $G$
  whether $G \in \cal{G}$ holds.
\end{corollary}

Nevertheless, an implicit precondition of the corollary is that the
set of all forbidden minors characterizing $\calG$ is known.

Another result by Robertson and Seymour~\cite{RS86a} is the 
so-called {\em grind minor theorem}, also known as the
{\em excluded grid theorem}.

\begin{theorem}[Grid minor theorem]\label{rs-grid}
  There is a function $f \colon \N \rightarrow \N$, such that every graph
with tree-width at least $f(k)$ has a $k \times k$-grid graph as minor.
\end{theorem}

Even if the grid minor theorem seems very technical
at first, it has a direct effect for the tree-width of a graph, as the
subsequent theorem from the same work by Robertson and Seymour~\cite{RS86a} shows.

\begin{theorem}\label{s-bw-m}
  For every planar graph $H$ there is an integer $k \in \N$, such that every
  graph without $H$ as minor has tree-width at most $k$.
\end{theorem}

Ahead of this result for tree-width, Robertson and Seymour~\cite{RS83}
proved an analogue result with respect to path-width.

\begin{theorem}\label{s-ww-m}
  For every forest $H$ there is an integer $k \in \N$, such that every
  graph without $H$ as minor has path-width at most $k$.
\end{theorem}

%%%%%%%%%%%%%%%%%%%%%%%%%%%%%%%%%%%%%%%%%%%%%%%%%%%%%%%%%%%%%%%%%%%%%%%%%%%
\subsection{Power of a Graph}\label{section-power}
For $d \in \N$ and a graph $G$, we denote the {\em $d$-th power} of
$G$ by $G^d$.
Thereby, $G^d$ possesses the same set of vertices as $G$, i.e.,
$V(G^d) = V(G)$, and two vertices of $G^d$ are adjacent if and only if
there exists a path of length at most $d$ in $G$ between these vertices.

To begin, the subsequent lemma provides an upper bound on the number
of neighbors each vertex might possess in the $d$-th power of a graph $G$.

\begin{lemma}\label{lm-gpowerd-neighbors}
  Given $d \in \N$ and a graph $G$, for every vertex $v \in V(G)$ it holds that
  \begin{align*}
    \deg_{G^d}(v) \leq \Delta(G) \sum_{i=0}^{d-1} (\Delta(G) - 1)^i.
  \end{align*}
\end{lemma}

\begin{proof}
  Let $v \in V(G)$ be a vertex of $G$.
  To obtain an upper bound on the number of neighbors $v$ might
  possess in $G^d$, we derive an upper bound for the number of vertices
  reachable from $v$ within a distance of at most $d$ in $G$.
  By definition, we know that $\deg_G(v) \leq \Delta(G)$ holds,
  such that $v$ has at most $\Delta(G)$ neighbors in $G$.
  By the same argument, every neighbor $u$ of $v$ in $G$ has at most
  $\Delta(G)$ neighbors.
  However, one of those neighbors of $u$ is $v$, such that $u$ has at
  most $\Delta(G) - 1$ not yet considered neighbors.
  Repeating the previous argument $d - 1$ times, we obtain an upper
  bound on the number of vertices with exact distance $d$ to $v$,
  \begin{align*}
    \Delta(G) (\Delta(G) - 1)^{d-1}.
  \end{align*}
  Since we are interested in an upper bound on the number of vertices
  with a distance of {\em at most} $d$ to $v$, we need to sum up
  the upper bounds for all intermediate distances from one up to $d$.
  Doing so, we obtain
  \begin{align*}
    \Delta(G) + \Delta(G)(\Delta(G) - 1) + \ldots +
    \Delta(G)(\Delta(G) - 1)^{d - 1} = \Delta(G)
    \sum_{i=0}^{d-1}(\Delta(G) - 1)^i.
  \end{align*}
  As every vertex, which is reachable from $v$ in $G$ within a distance of at
  most $d$, is adjacent to $v$ in $G^d$, it follows that
  \begin{align*}
     \deg_{G^d}(v) \leq \Delta(G) \sum_{i=0}^{d-1} (\Delta(G) - 1)^i
  \end{align*}
  holds.
\end{proof}

Note that the previously stated upper bound is tight.
To see this, consider a tree $T$ with root $v \in V(T)$ and three
binary subtrees of equal, but arbitrary depth adjacent to $v$.

Having this upper bound for a vertex's number neighbors in the
$d$-th power of a graph at hand, the subsequent theorem makes use of
it to study the effect of raising a graph to the power of $d$ with
respect to tree-width and path-width.

\begin{theorem}
  For $d \in \N$ and a graph $G$ it holds that
  \begin{align*}
    \tws(G) & \leq \tws(G^d) \leq (\tws(G) + 1) \left( 1 + \Delta(G) \sum_{i=0}^{d-1} (\Delta(G) - 1)^i
    \right)- 1 \quad
              \text{and} \\
    \pws(G) & \leq \pws(G^d) \leq (\pws(G) + 1) \left( 1 + \Delta(G) \sum_{i=0}^{d-1} (\Delta(G) -
    1)^i \right) - 1.
  \end{align*}
\end{theorem}

\begin{proof}
  The lower bounds follow by Corollary~\ref{cor-pw-tw-subgr}, as $G$ is a
  subgraph of $G^d$.

  For the upper bounds, denote by $(\calX, T)$ a tree-decomposition
  (by $\calX$ a path-decomposition) of $G$.
  Constructing $G^d$, we know all neighbors of every vertex $v
  \in V(G)$ in $G^d$.
  To obtain a tree-decomposition (path-decomposition) for $G^d$, for
  every vertex $v \in V(G)$, we add all neighbors of $v$ in $G^d$, i.e.,
  $N_{G^d}(v)$, to every bag $X \in \calX$ containing $v$.
  We denote the resulting bag by $X'$ and the resulting set of bags by
  $\calX'$.
  Afterwards, by Lemma~\ref{lm-gpowerd-neighbors}, every bag of
  $\calX'$ contains at most
  \begin{align*}
    (\tws(G) + 1) \left( 1 + \Delta(G) \sum_{i=0}^{d-1} (\Delta(G) - 1)^i
    \right) \\
    \left((\pws(G) + 1) \left( 1 + \Delta(G) \sum_{i=0}^{d-1} (\Delta(G) -
    1)^i \right)\right)
  \end{align*}
  vertices.
  Taking $G$ to the power of $d$ does not alter the set of
  vertices, i.e., $V(G) = V(G^d)$, so that $(\calX', T)$ ($\calX'$) still
  satisfies (tw-1) ((pw-1)).
  Adding, for every vertex $v \in V(G)$, all neighbors of $v$ in $G^d$
  to every bag containing $v$, all new edges introduced to $G^d$ are
  covered, such that $(\calX', T)$ ($\calX'$) satisfies (tw-2)
  ((pw-2)).
  To see that $(\calX', T)$ ($\calX'$) satisfies (tw-3) ((pw-3)), let
  $v \in V(G)$ be any vertex of $G$ and denote by $X_s', X_t' \in
  \calX'$ two bags with $v \in X_s', X_t'$.
  We separate three cases:
  (i) If it holds that $v \in X_s, X_t$ for $X_s, X_t \in \calX$, it
  immediately follows that $v$ is also in all bags connecting $X_s'$ and
  $X_t'$, since $(\calX, T)$ ($\calX$) satisfies (tw-3) ((pw-3)).
  (ii) If it holds that $v \in X_s$ but $v \not \in X_t$, there must be a
  vertex $u \in X_t$ with $\dist_G(u,v) \leq d$, as otherwise $v \in X_t'$
  would not hold.
  Since $\dist_G(u,v)\leq d$, there is a path $p_G$ between $u$ and
  $v$ in $G$ of length at most $d$.
  Consequently, with $(\calX, T)$
  ($\calX$) satisfying (tw-2) ((pw-2)) and (tw-3) ((pw-3)), there must
  also be a path $p_T$ between $s$ and
  $t$, such that for every $t'$ in $p_T$ there is a vertex $v'$ in
  $p_G$ with $v'\in X_{t'}$.
  For every such vertex $v'$ it holds that $\dist_G(v',v)\leq
  \dist_G(u,v)\leq d$, so that by the construction of $\calX'$ it
  follows that $v \in X'_{t'}$ holds.
  Therefore, also in this case all bags between $X_s'$ and $X_t'$
  must contain $v$.
  (iii) Finally, if $v \not \in X_s$ and $v \not \in X_t$, there must be at
  least one bag $X_r \in \calX$ with $v \in X_r$ and we can repeat the
  previous argument for $X_r', X_s'$ and $X_r', X_t'$.
  Consequently, $(\calX', T)$ ($\calX')$ is a tree-decomposition
  (path-decomposition) for $G^d$ of width
  \begin{align*}
    (\tws(G) + 1) \left( 1 +  \Delta(G) \sum_{i=0}^{d-1} (\Delta(G) - 1)^i
    \right)- 1 \\
    \left((\pws(G) + 1) \left( 1 + \Delta(G) \sum_{i=0}^{d-1} (\Delta(G) -
    1)^i \right) - 1 \right)
  \end{align*}
  and the upper bound follows.
\end{proof}

%%%%%%%%%%%%%%%%%%%%%%%%%%%%%%%%%%%%%%%%%%%%%%%%%%%%%%%%%%%%%%%%%%%%%%%%%%%
\subsection{Line Graph}\label{section-line}
In this subsection, we study the graph operation of creating a graph's
line graph, using the notation as formulated by Harary and
Norman~\cite{HN60}.
For a graph $G$, its {\em line graph}, $L(G)$, is defined by
\begin{align*}
  V(L(G)) & = \{ x_{\{ u, v \}}\mid \{ u, v \} \in E(G) \} \quad
            \text{and} \\
  E(L(G)) & = \{ \{ x_{e}, x_{f} \} \mid | e \cap f | = 1 \}.
\end{align*}
In other words, the line graph $L(G)$ of a graph $G$ has a vertex for every
edge of $G$ and two vertices of $L(G)$ are adjacent, if the corresponding
edges in $G$ are adjacent. The concept of a line graph, although not explicitly
called line graph,  was first used by Whitney~\cite{Whi32} in 1932.

Furthermore, the underlying graph $G$ of a given line graph $L(G)$ is
called the {\em root graph} of $L(G)$.

Given a graph $G$, it is possible to bound the tree-width (path-width)
of its line graph in the tree-width (path-width) of $G$ and $G$'s
maximum vertex degree.

\begin{theorem}\label{th-pw-tw-l}
  For a graph $G$ it holds that
  \begin{align*}
    \tws(G) - 1 & \leq \tws(L(G)) \leq (\tws(G) + 1) \Delta(G) - 1
                  \quad \text{and} \\
    \frac{1}{2} (\pws(G) + 1) - 1 & \leq \pws(L(G)) \leq (\pws(G) + 1)
                                    \Delta(G) - 1.
  \end{align*}
\end{theorem}

\begin{proof}
  The stated lower bound for the tree-width of $L(G)$ was shown by
  Harvey and Wood~\cite{HW18} in Proposition 2.3.

  A slightly weaker bound can be obtained as follows.
  Let $(\calX, T)$ be a tree-decomposition for $L(G)$ of width
  $\tws(L(G))$.
  In every bag of $(\calX, T)$ replace each edge of $G$ by both of its
  endpoints.
  Then, we can obtain a tree-decomposition for $G$ of width at most $2
  (\tws(L(G)) + 1) - 1$, such that $\frac{1}{2} (\tws(G) + 1) - 1 \leq
  \tws(L(G))$ follows.
  A similar argument results in the stated lower bound for path-width.

  The upper bounds are known from several works~\cite{Ats08, Bod93b,
    CFR03} and can be obtained as follows.
  Let $(\calX, T)$ be a tree-decomposition ($\calX$ a
  path-decomposition) for $G$ of width $\tws(G)$ ($\pws(G)$).
  If we replace every bag $X_u \in \calX$ by the set of all edges
  incident to at least one vertex in $X_u$, we obtain a
  tree-decomposition (path-decomposition) for $L(G)$ of width at most
  $(\tws(G) + 1) \Delta(G) - 1$ ($(\pws(G) + 1)\Delta(G) - 1$).
\end{proof}

Stricter upper bounds than the ones shown in Theorem~\ref{th-pw-tw-l}
can be found in Theorem 1.3 of the work by Harvey and
Wood~\cite{HW18}.

Furthermore, it is easy to confirm that for every graph $G$, the edges
incident to a vertex of $G$ with degree $\Delta(G)$ form a clique of
size $\Delta(G)$ in $L(G)$.
With Lemma~\ref{le-tw-clique} and Lemma~\ref{le-pw-clique}, the
subsequent corollary follows immediately.

\begin{corollary}\label{cor-tw-pw-line}
  For a graph $G$ it holds that $\tws(L(G)) \geq \Delta(G) - 1$
  and $\pws(L(G)) \geq \Delta(G) - 1$.
\end{corollary}

For special graphs the inequality turns into an equality, as the
subsequent result shows.

\begin{proposition}\label{th-tree}
  Let $G$ be a graph.
  If $\tws(G) = 1$ is true, it holds that $\tws(L(G)) =  \Delta(G) -
  1$.
  If $\pws(G) = 1$ is true, it holds that $\pws(L(G)) = \Delta(G) -
  1$.
\end{proposition}

The statement for tree-width was shown by Harvey and Wood~\cite{HW18}.
The statement for path-width follows by two arguments.
For the lower bound we refer to Corollary~\ref{cor-tw-pw-line}.
For the upper bound, we know from Section~\ref{subsection-minor} that
a graph $G$ of path-width one can be identified as disjoint union of
caterpillar graphs, allowing to construct a path-decomposition of
width at most $\Delta(G) - 1$ for $L(G)$.

Beside providing a direct equation how to obtain a line graph's
tree-width (path-width) from its root graph's tree-width (path-width),
if the root graph is a forest (caterpillar graph), the previous
proposition shows that a line graph's tree-width (path-width) can not
be bounded in its root graph's tree-width (path-width).

We conclude this subsection with a theorem by Harvey and
Wood~\cite{HW15}, showing how to derive a line graph's tree-width (path-width)
from its root graph, if the root graph is a complete graph.

\begin{theorem}
  For $n \geq 2$ it holds that
  \begin{align*}
    \tws(L(K_n)) = \pws(L(K_n)) = \begin{cases}
      \left( \frac{n-1}{2} \right) \left( \frac{n-1}{2} \right) + n
      -2, & \text{ if } n \text{ is odd,} \\
      \left( \frac{n-2}{2} \right) \left(\frac{n}{2}\right ) + n - 2,
          & \text{ if } n \text{ is even.}
    \end{cases}
  \end{align*}
\end{theorem}

%%%%%%%%%%%%%%%%%%%%%%%%%%%%%%%%%%%%%%%%%%%%%%%%%%%%%%%%%%%%%%%%%%%%%%%%%%%
\subsection{Edge Complement}\label{EC}
In this subsection of our work, we study the graph operation of
creating a graph's edge complement graph.
Thereby, the {\em edge complement graph} of a graph $G$, denoted by
$\co G$, has the same vertex set as $G$ and two vertices are adjacent
in $\co G$ if and only if they are not adjacent in $G$, i.e.,
\begin{align*}
  V(\co G) & = V(G) \quad \text{and} \\
  E(\co G) & =  \{ \{u, v\} \mid u,v \in V(G), u \neq v, \{u, v\}
  \not\in E(G) \}.
\end{align*}

Let $K_{1, \ell}$ be a star graph with one dominating vertex $v$ in
the center and $\ell$ vertices $u_i$, $1 \leq i \leq \ell$, as
satellites, all only adjacent to $v$.
Then, the edge complement graph of $K_{1, \ell}$, $\co K_{1, \ell}$,
consists of an isolated vertex $v$ and a clique of size $\ell$ formed
by all satellites $u_i$, $1 \leq i \leq \ell$.
Since neither $K_3$ nor $I(K_{1, 3})$ is a minor of $K_{1, \ell}$, we
know from Subsection~\ref{subsection-minor} that $\pws(K_{1, \ell}) =
\tws(K_{1, \ell}) = 1$ holds.
However, by Lemma~\ref{le-tw-clique} and Lemma~\ref{le-pw-clique} it
follows that $\tws(\co K_{1, \ell}) = \pws(\co K_{1, \ell}) = \ell -
1$ is true.
Therefore, it is generally impossible to bound the tree-width
(path-width) of an edge complement graph $\co G$ in the tree-width
(path-width) of the original graph $G$.

Nevertheless, Joret and Wood~\cite{JW12} proved the subsequent theorem, 
providing a lower bound for the
tree-width of a graph's edge complement graph.\footnote{Formulating
  bounds of the form $f(G) + f(\co G)$ for a graph parameter $f$ is
  known as {\em Nordhaus-Gaddum problem}.}

\begin{theorem}\label{tw-ng}
  For a graph $G$ it holds that
  \begin{align*}
    \tws(G) + \tws(\co G) \geq |V(G)| - 2.
  \end{align*}
\end{theorem}

In their work, Joret and Wood
also showed that the specified bound is tight.
As $\pws(G) \geq \tws(G)$ is true for every graph $G$, the subsequent
corollary follows immediately.

\begin{corollary}\label{pw-ng}
  For a graph $G$ it holds that
  \begin{align*}
    \pws(G) + \pws(\co G) \geq |V(G)| - 2.
  \end{align*}
\end{corollary}

For the path with four vertices, $P_4$, it holds that $\co P_4 \cong
P_4$ as well as $\pws(P_4) = 1$.
Consequently, we obtain
\begin{align*}
  \pws(P_4) + \pws(\co P_4) = \pws(P_4) + \pws(P_4) = 2 = |V(P_4)| - 2,
\end{align*}
such that the bound specified in Corollary~\ref{pw-ng} is tight, too.

%%%%%%%%%%%%%%%%%%%%%%%%%%%%%%%%%%%%%%%%%%%%%%%%%%%%%%%%%%%%%%%%%%%%%%%%%%%
\subsection{Local Complementation}
In his work, Bouchet~\cite{Bou94} introduced the graph operation local
complementation.
Given a graph $G$ and a vertex $v \in V(G)$, the {\em local
complementation} of $G$, denoted by $LC(G,v)$, is defined by
\begin{align*}
  V(LC(G,v)) = V(G) & \quad \text{and} \\
  E(LC(G,v)) = E(G) & \setminus \{ \{u, w\} \mid u,w \in N_G(v) \} \\
                    & \cup \{ \{u, w\} \mid u,w \in N_G(v), u \neq w,
                      \{u,w\}\not \in E(G) \}.
\end{align*}

In other words, $LC(G,v)$ is obtained from graph $G$ by replacing the
subgraph of $G$ induced by $N_G(v)$ with its edge complement.
Recall that $v \not \in N_G(v)$ holds, such that the neighborhood of
$v$ in $LC(G,v)$ is the same as in $G$.

Denote by $K_{1,\ell}$ the star graph we made already use of in
Subsection~\ref{EC}.
Applying a local complementation to the dominating vertex $v$, it is
easy to see that $LC(K_{1,\ell}, v)$ equals a clique of size $\ell +
1$.
The star $K_{1,\ell}$ has tree-width and path-width one, while
$LC(K_{1,\ell}, v)$ has tree-width and path-width $\ell$.
Therefore, in general the tree-width (path-width) of a graph $G$'s
local complement $LC(G,v)$ can not be bounded in the tree-width
(path-width) of $G$.

%%%%%%%%%%%%%%%%%%%%%%%%%%%%%%%%%%%%%%%%%%%%%%%%%%%%%%%%%%%%%%%%%%%%%%%%%%%
\subsection{Seidel Switching}
The Seidel switching operation was introduced by the Dutch
mathematician J.~J.~Seidel in connection with regular structures, such
as systems of equiangular lines, strongly regular graphs, or so-called
two-graphs~\cite{Sei74a,Sei76,ST81}.
Several examples for applications of Seidel switching can be found in
algorithms, e.g., in a polynomial-time algorithm for the
$P_3$-structure recognition problem~\cite{Hay96} or in an algorithm
for the construction of bi-join decompositions of graphs~\cite{MR05}.

For a graph $G$ and a vertex $v \in V(G)$, the graph resulting from a
{\em Seidel switching} operation, denoted by $S(G,v)$, is defined as
follows.
The vertex set of $S(G,v)$ is the same as the vertex set of $G$, i.e.,
$V(S(G,v)) = V(G)$, and the edge set of $S(G, v)$ is defined as
\begin{align*}
  E(S(G,v)) = E(G) & \setminus \{\{v,w\} \mid w \in N_G(v) \} \\
                   & \cup \{\{v,w\} \mid w \in V(G) \setminus (N_G(v)
                     \cup \{ v \}) \}.
\end{align*}
In other words, every neighbor of $v$ in $G$ is a non-neighbor of $v$
in $S(G,v)$ and every non-neighbor of $v$ in $G$ is a neighbor of $v$
in $S(G,v)$.

Given this definition of Seidel switching, subsequently, extending a
result by Bodlaender and Hage~\cite{BH12}, we show that a single Seidel switching operation increases or
decreases a graph's tree-width and path-width at most by one.

\begin{theorem}\label{Tswitch}
  For a graph $G$ and a vertex $v \in V(G)$ it holds that
  \begin{align*}
    \tws(G) - 1 & \le \tws(S(G,v)) \le \tws(G) + 1 \quad \text{and} \\
    \pws(G) - 1 & \le \pws(S(G,v)) \le \pws(G) + 1.
  \end{align*}
\end{theorem}

\begin{proof}
  For the upper bounds, let $(\calX, T)$ be a tree-decomposition
  ($\calX$ a path-decomposition) for $G$ of width $\tws(G)$
  ($\pws(G)$).
  When we add $v$ to all bags of $\calX$, denoting the resulting set
  of bags by $\calX'$, we obtain a tree-decomposition $(\calX', T)$ (a
  path-decomposition $\calX'$) for $S(G, v)$ of width at most $\tws(G)
  + 1$ ($\pws(G) + 1$).
  Consequently, $\tws(S(G,v)) \leq \tws(G) + 1$ ($\pws(S(G,v)) \leq
  \pws(G) + 1$) follows.

  Since $S(S(G, v), v) = G$ holds, we can derive the lower bound from
  the upper bound via $\tws(G) = \tws(S(S(G, v), v) \leq \tws(S(G,v))
  + 1$ ($\pws(G) = \pws(S(S(G,v),v)) \leq \pws(S(G,v)) + 1$).
\end{proof}

Note that the bounds shown in Theorem~\ref{Tswitch} are tight.
To convince oneself of this fact, consider the path with five
vertices, $P_5$.
Its tree-width and path-width is one.
Denote by $v$ one of the two vertices in $P_5$ with degree one.
Then, $S(P_5,v)$ contains $K_3$ as minor, such that a tree-width and
path-width of at least two follows.
Following the example in the opposite direction, i.e., applying the
Seidel switching operation to $S(P_5, v)$ for the same vertex $v$,
provides an example that the lower bound is tight, too.

In 1980, Colbourn and Corneil~\cite{CC80} studied the
complexity of the decision problem whether two graphs are switching
equivalent.
In their work, they proved that this decision problem is polynomial
time equivalent to the decision problem of graph isomorphism.
Thereby, two graphs $G$ and $G'$ with the same vertex set $V$ are
called {\em switching equivalent}, if there exists a sequence of
vertices $(v_1, \ldots, v_\ell)$ in $V$, such that for $G^0 = G$ and
$G^i = S(G^{i-1},v_i)$, $1 \leq i \leq \ell$, it holds that $G^\ell =
G'$.
In 2012, Bodlaender and Hage~\cite{BH12} considered in their work the
tree-width of switching classes.
By the definition of switching equivalence and via
Theorem~\ref{Tswitch}, we can formulate the subsequent corollary,
contributing to the research on the tree-width of switching classes,
initiated by Bodlaender and Hage.

\begin{corollary} \label{the-sw}
  Let $G$ and $G'$ be two switching equivalent graph and denote by
  $(v_1, \ldots, v_\ell)$ a sequence of vertices, such that $G^{\ell}
  = G'$ is true. Then, it holds that
  \begin{align*}
    \tws(G') \le \tws(G) + \ell \quad \text{and} \quad \pws(G') \le
    \pws(G) + \ell.
  \end{align*}
\end{corollary}

%%%%%%%%%%%%%%%%%%%%%%%%%%%%%%%%%%%%%%%%%%%%%%%%%%%%%%%%%%%%%%%%%%%%%%%%%%%
\subsection{Seidel Complementation}\label{sec-sc}
Limouzy~\cite{Lim10} defined the Seidel complementation operation in order to give
a characterization for permutation graphs.
For a graph $G$ and a vertex $v \in V(G)$, the graph resulting from
the {\em Seidel complementation} operation, denoted by $SC(G,v)$, has
the same vertices as $G$, i.e., $V(SC(G,v)) = V(G)$, and edge set
\begin{align*}
  E(SC(G,v)) = E(G) \triangle \{\{x, y\} \mid \{v,x\} \in E(G),
  \{v,y\} \not\in E(G) \}.
\end{align*}
In other words, the edge set of $SC(G,v)$ equals the edge set of $G$
with edges and non-edges between the neighborhood and non-neighborhood
of $v$ complemented.

Let $G$ be a graph that consists of two parts.
The first part is a star with vertex $v$ as dominating vertex
and $\ell \in \N$ satellites $u_1, \ldots, u_{\ell}$, adjacent to $v$.
The second part is a
set of $\ell$ isolated vertices, $w_1, \ldots, w_{\ell}$.
Since $G$ does neither possess a $K_3$ nor an $I(K_{1,3})$ as minor, we
know that $\tws(G) = \pws(G) = 1$ holds.
Applying the Seidel complementation operation to vertex $v$, $SC(G,v)$
contains a complete, bipartite subgraph formed by the vertices $u_1,
\ldots, u_{\ell}, w_1, \ldots, w_{\ell}$.
By Lemma~\ref{le-tw-bip-clique}
(Lemma~\ref{le-pw-bip-clique}) it follows that $\tws(SC(G,v)) \geq
\ell - 1$ ($\pws(SC(G,v)) \geq \ell - 1$) is true.
Consequently, we conclude that, given a graph $G$ and a vertex $v \in
V(G)$, the tree-width (path-width) of $SC(G,v)$ cannot be bounded in
the tree-width (path-width) of $G$.

%%%%%%%%%%%%%%%%%%%%%%%%%%%%%%%%%%%%%%%%%%%%%%%%%%%%%%%%%%%%%%%%%%%%%%%%%%%
%%%%%%%%%%%%%%%%%%%%%%%%%%%%%%%%%%%%%%%%%%%%%%%%%%%%%%%%%%%%%%%%%%%%%%%%%%%
%%%%%%%%%%%%%%%%%%%%%%%%%%%%%%%%%%%%%%%%%%%%%%%%%%%%%%%%%%%%%%%%%%%%%%%%%%%
\section{Binary Graph Operations}\label{section-binary-operations}
Let $G_1$, $G_2$ be two non-empty graphs and $f$ a binary graph
operation that creates a new graph $f(G_1, G_2)$ from $G_1$ and $G_2$.
In this section, we consider the tree-width and path-width of $f(G_1,
G_2)$ with respect to the tree-widths and path-widths of the initial
graphs $G_1$ and $G_2$.
In particular, we study the binary graph operations disjoint union,
join, union, substitution, various types of graph products, 1-sum, and
corona.

%%%%%%%%%%%%%%%%%%%%%%%%%%%%%%%%%%%%%%%%%%%%%%%%%%%%%%%%%%%%%%%%%%%%%%%%%%%
\subsection{Disjoint Union}\label{section-disj-union}
The {\em disjoint union} of two vertex-disjoint graphs $G_1$ and
$G_2$, denoted by $G_1 \oplus G_2$, is defined as the graph with
vertex set $V(G_1) \cup V(G_2)$ and edge set $E(G_1) \cup E(G_2)$.

Bodlaender and M\"ohring~\cite{BM90,BM93} proved the subsequent theorem
with respect to the tree-width and path-width of a graph which is the
disjoint union of two vertex-disjoint graphs.

\begin{theorem}\label{th-du}
  Let $G_1$ and $G_2$ be two vertex-disjoint graphs, then it holds
  that ${\tws(G_1 \oplus G_2)=\max(\tws(G_1),\tws(G_2))}$ and $\pws(G_1
  \oplus G_2)=\max(\pws(G_1),\pws(G_2))$.
\end{theorem}

These bounds imply that the tree-width and path-width of a graph can
be derived from the tree-width and path-width of its connected
components.

\begin{corollary}
  Let $G$ be a graph.
  It holds that
  \begin{enumerate}
  \item
    the tree-width of $G$ is the maximum tree-width and
  \item
    the path-width of $G$ is the maximum path-width
  \end{enumerate}
  of its connected components.
\end{corollary}

%%%%%%%%%%%%%%%%%%%%%%%%%%%%%%%%%%%%%%%%%%%%%%%%%%%%%%%%%%%%%%%%%%%%%%%%%%%
\subsection{Join}\label{section-join}
The {\em join} of two vertex-disjoint graphs $G_1$ and $G_2$, denoted
by $G_1 \otimes G_2$, is defined as the graph with vertex set $V(G_1) \cup
V(G_2)$ and edge set
\begin{align*}
  E(G_1) \cup E(G_2) \cup \{ \{v_1, v_2\} \mid v_1\in V(G_1), v_2\in
  V(G_2) \}.
\end{align*}

As for the disjoint union of two graphs, Bodlaender and
M\"ohring~\cite{BM90,BM93} proved the subsequent theorem with respect to
the tree-width and path-width of a graph that is the join of two
vertex-disjoint graphs.

\begin{theorem}\label{th-jo}
  Let $G_1$ and $G_2$ be two vertex-disjoint graphs.
  Then, for the join of $G_1$ and $G_2$ it holds that
  ${\tws(G_1\otimes
    G_2)=\min(\tws(G_1)+|V(G_2)|, \tws(G_2)+|V(G_1)|)}$ and
  ${\pws(G_1\otimes
    G_2)=\min(\pws(G_1)+|V(G_2)|,\pws(G_2)+|V(G_1)|)}$.
\end{theorem}

The combination of Theorem~\ref{th-du} and~\ref{th-jo} implies that
for every co-graph $G$, it holds that $\tws(G) = \pws(G)$ and both
widths can be computed in linear time~\cite{BM90,BM93}.

%%%%%%%%%%%%%%%%%%%%%%%%%%%%%%%%%%%%%%%%%%%%%%%%%%%%%%%%%%%%%%%%%%%%%%%%%%%
\subsection{Union}\label{sec-sum}
The {\em union} of two graphs $G_1$ and $G_2$ with $V(G_1) =
V(G_2)$, denoted by $G_1 \cup G_2$, is defined as the graph with
vertices $V(G_1)$ and edge set $E(G_1) \cup E(G_2)$.
Thus, two vertices are adjacent in $G_1 \cup G_2$ if and only if they
are adjacent in $G_1$ or in $G_2$.

In general, it is not possible to bound the tree-width (path-width)
of the union of two graphs in terms of the tree-widths (path-widths) of
the individual graphs.
To see why that is the case, consider the subsequent example.

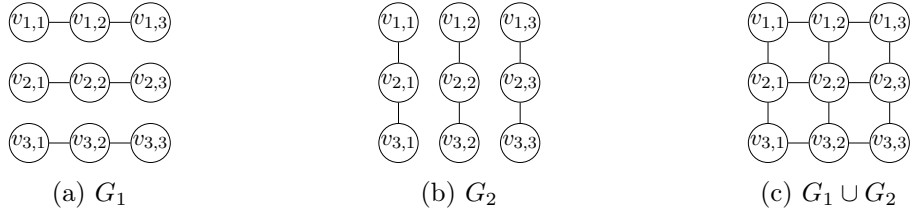
\begin{figure}
  \centering
  \begin{subfigure}{0.30\textwidth}
    \centering
    \scalebox{0.8}{
      \begin{tikzpicture}
        \draw (-1,1) \gnode (v11) {$v_{1,1}$};
        \draw (0,1) \gnode (v12) {$v_{1,2}$};
        \draw (1,1) \gnode (v13) {$v_{1,3}$};
        \draw (-1,0) \gnode (v21) {$v_{2,1}$};
        \draw (0,0) \gnode (v22) {$v_{2,2}$};
        \draw (1,0) \gnode (v23) {$v_{2,3}$};
        \draw (-1,-1) \gnode (v31) {$v_{3,1}$};
        \draw (0,-1) \gnode (v32) {$v_{3,2}$};
        \draw (1,-1) \gnode (v33) {$v_{3,3}$};
        \draw[-] (v11) to (v12);
        \draw[-] (v12) to (v13);
        \draw[-] (v21) to (v22);
        \draw[-] (v22) to (v23);
        \draw[-] (v31) to (v32);
        \draw[-] (v32) to (v33);
      \end{tikzpicture}
    }
    \caption{$G_1$}
  \end{subfigure}
  \hspace{1ex}
  \begin{subfigure}{0.30\textwidth}
    \centering
    \scalebox{0.8}{
      \begin{tikzpicture}
        \draw (-1,1) \gnode (v11) {$v_{1,1}$};
        \draw (0,1) \gnode (v12) {$v_{1,2}$};
        \draw (1,1) \gnode (v13) {$v_{1,3}$};
        \draw (-1,0) \gnode (v21) {$v_{2,1}$};
        \draw (0,0) \gnode (v22) {$v_{2,2}$};
        \draw (1,0) \gnode (v23) {$v_{2,3}$};
        \draw (-1,-1) \gnode (v31) {$v_{3,1}$};
        \draw (0,-1) \gnode (v32) {$v_{3,2}$};
        \draw (1,-1) \gnode (v33) {$v_{3,3}$};
        \draw[-] (v11) to (v21);
        \draw[-] (v21) to (v31);
        \draw[-] (v12) to (v22);
        \draw[-] (v22) to (v32);
        \draw[-] (v13) to (v23);
        \draw[-] (v23) to (v33);
      \end{tikzpicture}
    }
    \caption{$G_2$}
  \end{subfigure}
  \hspace{1ex}
  \begin{subfigure}{0.30\textwidth}
    \centering
    \scalebox{0.8}{
      \begin{tikzpicture}
        \draw (-1,1) \gnode (v11) {$v_{1,1}$};
        \draw (0,1) \gnode (v12) {$v_{1,2}$};
        \draw (1,1) \gnode (v13) {$v_{1,3}$};
        \draw (-1,0) \gnode (v21) {$v_{2,1}$};
        \draw (0,0) \gnode (v22) {$v_{2,2}$};
        \draw (1,0) \gnode (v23) {$v_{2,3}$};
        \draw (-1,-1) \gnode (v31) {$v_{3,1}$};
        \draw (0,-1) \gnode (v32) {$v_{3,2}$};
        \draw (1,-1) \gnode (v33) {$v_{3,3}$};
        \draw[-] (v11) to (v12);
        \draw[-] (v12) to (v13);
        \draw[-] (v21) to (v22);
        \draw[-] (v22) to (v23);
        \draw[-] (v31) to (v32);
        \draw[-] (v32) to (v33);
        \draw[-] (v11) to (v21);
        \draw[-] (v21) to (v31);
        \draw[-] (v12) to (v22);
        \draw[-] (v22) to (v32);
        \draw[-] (v13) to (v23);
        \draw[-] (v23) to (v33);
      \end{tikzpicture}
    }
    \caption{$G_1 \cup G_2$}
  \end{subfigure}
  \caption{$G_1$, $G_2$, and $G_1 \cup G_2$ for $m = n = 3$ in
    Example~\ref{ex-union-grid}.}
  \label{F-ex-union}
\end{figure}

\begin{example}\label{ex-union-grid}
  For $m, n \in \N$, define the set of vertices $V = \{ v_{i,j} \mid 1
  \leq i \leq m, 1 \leq j \leq n \}$.
  Next, denote by $G_1$ the disjoint union of $m$ paths
  with $n$ vertices from $V$, $P_n^i = (\{ v_{i, j} \mid 1 \leq j \leq n \},
  \{ \{ v_{i,j}, v_{i, j+1} \} \mid 1 \leq j < n \})$, $1 \leq i \leq
  m$, and by $G_2$ the disjoint union of $n$ paths with $m$ vertices
  from $V$, $P_m^j = (\{ v_{i,j} \mid 1 \leq i \leq m \}, \{ \{
  v_{i,j}, v_{i+1,j} \} \mid 1 \leq i < m \})$, $1 \leq j \leq n$.
  Since paths have a tree-width (path-width) of $1$, it follows by
  Theorem~\ref{th-du} that $\tws(G_1) = \tws(G_2) = 1$
  ($\pws(G_1) = \pws(G_2) = 1$) holds.

  The union, $G_1 \cup G_2$, of $G_1$ and $G_2$ is an $n \times
  m$-grid graph, and Bodlaender~\cite{Bod98} proved that the tree-width 
  (path-width) of an $n \times m$-grid graph equals $\min(n, m)$.
  Consequently, it is not possible to bound the tree-width
  (path-width) of $G_1 \cup G_2$ in the tree-widths (path-widths) of
  $G_1$ and $G_2$.

  See Figure~\ref{F-ex-union} for an explicit example of $G_1, G_2$,
  and the resulting union $G_1 \cup G_2$.
\end{example}

%%%%%%%%%%%%%%%%%%%%%%%%%%%%%%%%%%%%%%%%%%%%%%%%%%%%%%%%%%%%%%%%%%%%%%%%%%%
\subsection{Substitution}\label{Substi-oper}
Let $G_1$ and $G_2$ be two vertex-disjoint graphs and $v \in V(G_1)$
a vertex.
The {\em substitution} of $v$ by $G_2$ in $G_1$, denoted by
$G_1[v/G_2]$, is defined as the graph with vertex set $(V(G_1)
\setminus \{ v \}) \cup V(G_2)$ and edge set
\begin{align*}
  (E(G_1) \setminus \{ \{v, w\} \mid w \in N_{G_1}(v) \})
  \cup \{ \{ u, w \} \mid u \in V(G_2), w \in N_{G_1}(v) \} \cup E(G_2).
\end{align*}
The vertex set $V(G_2)$ is called a {\em module} of graph
$G_1[v/G_2]$, since all vertices of $V(G_2)$ are adjacent to the same
vertices of $V(G_1[v/G_2]) \setminus V(G_2)$ in $G_1[v/G_2]$.

Given this definition of substitution, the subsequent theorem
considers the tree-width (path-width) of $G_1[v/G_2]$ in terms of the
tree-width (path-width) of $G_1$ and $G_2$.

\begin{theorem}\label{th-substitution}
  For two vertex-disjoint graphs $G_1$, $G_2$ and vertex $v \in
  V(G_1)$ it holds that
  \begin{align*}
    \max(\tws(G_1),\tws(G_2)) \leq \tws(G_1[v/G_2]) \leq
    \min(\tws(G_1) + |V(G_2)|, \tws(G_2) + |V(G_1)|) - 1
  \end{align*}
  and
  \begin{align*}
    \max(\pws(G_1),\pws(G_2)) \leq \pws(G_1[v/G_2])\leq
    \min(\pws(G_1) + |V(G_2)|, \pws(G_2)+|V(G_1)|) - 1.
  \end{align*}
\end{theorem}

\begin{proof}
  The lower bounds follow by Corollary~\ref{cor-pw-tw-subgr}, as $G_1$
  and $G_2$ are subgraphs of $G_1[v/G_2]$.
  For the upper bounds, we can
  \begin{enumerate}
  \item
    replace $v$ in every bag of a tree-decomposition
    (path-decomposition) of width $\tws(G_1)$ ($\pws(G_1)$) for $G_1$ by
    $V(G_2)$ or
  \item
    add to every bag of a tree-decomposition (path-decomposition) of
    width $\tws(G_2)$ ($\pws(G_2)$) for $G_2$ the set $V(G_1) \setminus
    \{v\}$.
  \end{enumerate}
  As both alternatives result in a valid tree-decomposition
  (path-decomposition) for $G_1[v/G_2]$, the upper bound follows.
\end{proof}

Note that the upper bounds described in the previous theorem are
strict, as for two cliques $K_n, K_m$ and a vertex $v \in V(K_n)$, we have
$K_n[v/K_m] = K_{n+m-1}$ with
\begin{align*}
  \tws(K_{n+m-1}) = n + m - 2 = \min(\tws(K_n) + m, \tws(K_m) + n) - 1
  \quad \text{and} \\
  \pws(K_{n+m-1}) = n + m - 2 = \min(\pws(K_n) + m, \pws(K_m) + n) - 1.
\end{align*}

Besides this general upper bound for tree-width, the ensuing
proposition can provide an even stricter upper bound in specific
situations.

\begin{proposition}
  Let $G_1$, $G_2$ be two vertex-disjoint graphs and $v \in V(G_1)$ a
  non-isolated vertex.
  Then, it holds that
  \begin{align*}
    \tws(G_1[v/G_2]) \leq \max( \tws(G_1) - 1, \tws(G_2) ) + |N_{G_1}(v)|.
  \end{align*}
\end{proposition}

\begin{proof}
  Let $(\calX_1, T_1)$ be a tree-decomposition for $G_1$ of width
  $\tws(G_1)$ and $(\calX_2, T_2)$ a tree-de\-com\-po\-sition for $G_2$ of
  width $\tws(G_2)$.
  Replace $v$ by $N_{G_1}(v)$ in every bag of $\calX_1$ and
  denote the modified set of bags by $\calX_1'$.
  Then, $(\calX_1', T_1)$ is a tree-decomposition for $G_1 \setminus
  \{v\}$ of width at most $\tws(G_1) - 1 + |N_{G_1}(v)|$.
  Subsequently, add $N_{G_1}(v)$ to all bags of $\calX_2$ and denote
  the modified set of bags by $\calX_2'$.
  Let $v_1 \in V(T_1)$ be a vertex of $T_1$ with $N_{G_1}(v) \subseteq
  X_{v_1}$ for $X_{v_1} \in \calX_1'$ and $v_2$ any vertex of $T_2$.
  Then, $(\calX, T)$ with $\calX = \calX_1' \cup \calX_2'$, $V(T) =
  V(T_1) \cup V(T_2)$, and $E(T) = E(T_1) \cup E(T_2) \cup \{ v_1, v_2
  \}$ is a tree-decomposition for $G_1[v/G_2]$ of width at most $\max(
  \tws(G_1) - 1, \tws(G_2)) + |N_{G_1}(v)|$.
\end{proof}

%%%%%%%%%%%%%%%%%%%%%%%%%%%%%%%%%%%%%%%%%%%%%%%%%%%%%%%%%%%%%%%%%%%%%%%%%%%
\subsection{Graph Product}
The graph product of two vertex-disjoint graphs $G_1$ and $G_2$ is a
new graph with vertex set $V(G_1) \times V(G_2)$ and an edge set
derived from the adjacency, equality, or non-adjacency of vertices in
the original graphs $G_1$ and $G_2$.
In this work, we consider the
cartesian~\cite{Sab60}, 
categorical~\cite{Wei62}, 
co-normal~\cite{HW67},
lexicographic~\cite{Har58}, and 
normal~\cite{Sab60} 
graph product, as well as the
symmetric difference~\cite{HW67} and 
the rejection~\cite{HW67}.\footnote{Weichsel introduced the categorical graph product as the
  ``Kronecker product'', while Harary and Wilcox referred to it as
  ``conjunction''.
  The normal graph product was introduced by Sabidussi as the ``strong
  product''.
  The co-normal graph product was introduced by Harary and Wilcox as
  ``disjunction'' and the lexicographic graph product was initially
  defined by Harary as ``composition''.}
All graph products, their respective notations, and the edge sets of
the resulting graphs are listed in Table~\ref{tab-gp}.
For more exhaustive definitions and in-depths results on these graph
products, we refer to the works by Imrich and Klavzar~\cite{IK00} and
Jensen and Toft~\cite{JT94}.

\begin{table}[ht]
  \centering
  \begin{tabular}{lll}
    \toprule
    Graph product        & Notation & Edge set $\{\{(u_1,u_2),(v_1,v_2)\} \mid$ \\
    \midrule
    Cartesian            & $G_1 \times G_2$ & $(u_1=v_1 \wedge \{u_2,v_2\}\in E(G_2))$ \\
                         & & \quad $\vee  (u_2=v_2 \wedge \{u_1,v_1\}\in E(G_1))\}$ \\
    Categorical          & $G_1 \catprod G_2$ & $\{u_1,v_1\} \in E(G_1)  \wedge \{u_2,v_2\} \in E(G_2)\}$ \\
    Co-Normal            & $G_1 \vee G_2$ & $\{u_1,v_1\} \in E(G_1)  \vee \{u_2,v_2\} \in E(G_2)\} $ \\
    Lexicographic        & $G_1[G_2]$ & $(\{u_1,v_1\} \in E(G_1)) \vee (u_1=v_1 \wedge \{u_2,v_2\} \in E(G_2))\}$ \\
    Normal (Strong)      & $G_1 \star G_2$ & $(u_1 = v_1 \wedge \{u_2,v_2\} \in E(G_2))$ \\
                         & & \quad $\vee (\{u_1,v_1\} \in E(G_1) \wedge u_2 = v_2)$ \\
                         & & \quad $\vee (\{u_1,v_1\} \in E(G_1) \wedge \{u_2,v_2\} \in E(G_2))\}$ \\
    Symmetric difference & $G_1 \triangle G_2$ & $\{u_1,v_1\} \in E(G_1) \veebar  \{u_2,v_2\} \in E(G_2)\}$ \\
    Rejection            & $G_1 | G_2$ & $\{u_1,v_1\} \not\in E(G_1) \wedge \{u_2,v_2\} \not\in E(G_2)\}$ \\
    \bottomrule
  \end{tabular}
  \caption{Graph products}\label{tab-gp}
\end{table}

Besides these graph products, Teh and Yap defined the $\gamma$-product of
two graphs $G_1$ and $G_2$ as $\overline{\overline{G_1} \vee
  \overline{G_2}}$~\cite{TY64}.
Transforming the edge set of the $\gamma$-product of two graphs
$G_1$ and $G_2$,
it follows that $\overline{\overline{G_1} \vee \overline{G_2}} = G_1
\star G_2$ holds.
In other words, the $\gamma$-product is merely a different formulation
for the normal product of two graphs.
Furthermore, transforming the edge set of $G_1 | G_2$,
we obtain $G_1 | G_2 = \overline{G_1} \star \overline{G_2}$ while
transforming the edge set of $G_1 \triangle G_2$
results in $G_1 \triangle G_2 = (G_1 \times G_2) \cup \left(G_1 \catprod
\overline{G_2} \right) \cup \left(\overline{G_1} \catprod G_2 \right)$.

For two paths $P_n$, $P_m$ with $n, m \in \N$ it holds that $P_n \times
P_m$ is an $n \times m$-grid graph.
Consequently, for two graphs $G_1$ and $G_2$, it is generally not
possible to bound the tree-width (path-width) of $G_1 \times G_2$ from
above by the tree-widths (path-widths) of $G_1$ and $G_2$, c.f.
Example~\ref{ex-union-grid}.
With $P_n \times P_m$ being a subgraph of $P_n \vee P_m$, $P_n \star
P_m$, and $P_n \triangle P_m$, the same result follows by
Corollary~\ref{cor-pw-tw-subgr} for the co-normal and normal graph
product as well as the symmetric difference.
Next, $P_n \catprod P_m$ has an $a \times a$-grid graph with $a$
proportional to $\min(n, m)$ as subgraph, so that by the same
argument as before, the tree-width (path-width) of $G_1 \catprod G_2$ can
generally not be bound from above by the tree-widths (path-widths) of
$G_1$ and $G_2$.
For $n, m \in \N$, let $I_n$ and $I_m$ denote the graphs that contain
$n$ and $m$ isolated vertices.
By our earlier observation, we know that $I_n | I_m = K_n \star K_m$
holds.
Consequently, the tree-width (path-width) of $I_n | I_m$ cannot be
bound from above by the tree-widths (path-widths) of $I_n$ and $I_m$,
as $\tws(I_n) = \pws(I_n) = 1$ holds, while $K_n$ and $K_m$ are
subgraphs of $I_n | I_m$, such that by Lemma~\ref{le-tw-clique}
(Lemma~\ref{le-pw-clique}) $\tws(I_n | I_m) \geq \max(n, m) - 1$
($\pws(I_n | I_m) \geq \max (n, m) - 1$) follows.
Therefore, for two graphs $G_1$ and $G_2$, the tree-width (path-width)
of the rejection $G_1 | G_2$ can in general not be bound from above
by the tree-widths (path-widths) of $G_1$ and $G_2$.
The subsequent corollary summarizes these observations.

\begin{corollary}
  Let $G_1$ and $G_2$ be two graphs.
  It is not possible to provide an upper bound for the tree-width
  (path-width) of the two graph's cartesian, categorical, co-normal, and
  normal graph product, their rejection or their symmetric difference in
  terms of the tree-widths (path-widths) of $G_1$ and $G_2$.
\end{corollary}

Lower bounds for the tree-width (using the notation of bramble
number~\cite{Ree97}) of the cartesian and the normal product of two
graphs are given in terms of Hadwiger, PI, and bramble number in the
work by by Kozawa, Otachi, and Yamazaki~\cite{KOY14}.

Having discussed all previously defined graph products but the
lexicographic graph product, subsequently we provide bounds for this
operation.
First, for two graphs $G_1$ and $G_2$, it holds that $G_1$ and $G_2$
are subgraphs of $G_1[G_2]$.
Consequently, by Corollary~\ref{cor-pw-tw-subgr} it follows that the
tree-width (path-width) of $G_1[G_2]$ is at least as big as the
maximum over the tree-widths (path-widths) of $G_1$ and $G_2$.
Second, to obtain a tree-decomposition (path-decomposition) for $G_1[G_2]$, we
begin with a tree-decomposition (path-decomposition) for $G_1$ and
replace every vertex $v_i \in V(G_1)$ in every bag by $(v_i, v_j)$ for
all $v_j \in V(G_2)$.
This results in a tree-decomposition (path-decomposition) for
$G_1[G_2]$ of width $(\tws(G_1) + 1) |V(G_2)| - 1$ ($(\pws(G_1) + 1)
|V(G_2)| - 1$).

\begin{corollary}\label{Tlex}
  Let $G_1$ and $G_2$ be two vertex-disjoint graphs.
  It holds that
  \begin{align*}
    \max(\tws(G_1), \tws(G_2)) & \leq \tws(G_1[G_2]) \leq (\tws(G_1) + 1)
                                 |V(G_2)| - 1 \quad \text{and},\\
    \max(\pws(G_1), \pws(G_2)) & \leq \pws(G_1[G_2]) \leq (\pws(G_1) +
                                 1) |V(G_2)| - 1.
  \end{align*}
\end{corollary}

Bodlaender et al.~\cite{BGHK95} have shown that if $G_2$ is a clique,
the upper bounds for $\tws(G_1[G_2])$ and $\pws(G_1[G_2])$ as stated
above are tight.

\begin{theorem}
  Let $G$ be a graph and $q \in \N$.
  It holds that $\tws(G[K_{q}]) = (\tws(G) + 1) q - 1$ and
  $\pws(G[K_{q}]) = (\pws(G) + 1) q - 1$.
\end{theorem}

%%%%%%%%%%%%%%%%%%%%%%%%%%%%%%%%%%%%%%%%%%%%%%%%%%%%%%%%%%%%%%%%%%%%%%%%%%%
\subsection{1-Sum}
Let $G_1$ and $G_2$ be two vertex-disjoint graphs and $v \in V(G_1)$
and $w \in V(G_2)$ two vertices.
The {\em 1-sum} of $G_1$ and $G_2$, denoted by $G_1 \oplus_{v,w} G_2$,
consists of the disjoint union of $G_1$ and $G_2$ with vertices $v$
and $w$ identified.
More specifically, graph $G_1 \oplus_{v,w} G_2$ has vertex set
$(V(G_1) \cup V(G_2) \cup \{z\}) \setminus \{v, w\} $ for a newly
introduced vertex $z$ and edge set
\begin{align*}
  E(G_1) \cup E(G_2)  \setminus & \{ \{v,v_1\} \mid v_1 \in N_{G_1}(v) \} \\
  \setminus & \{ \{w,w_1\} \mid w_1 \in N_{G_2}(w) \} \\
  \cup & \{ \{z, z_1\}\mid z_1 \in (N_{G_1}(v) \cup N_{G_2}(w)) \}.
\end{align*}

Having the 1-sum of two graphs $G_1$ and $G_2$ formally defined, in
the subsequent theorem we consider the tree-width and path-width of
$G_1 \oplus_{v,w} G_2$.

\begin{theorem}\label{T1s}
  Let $G_1$ and $G_2$ be two vertex-disjoint graphs and $v \in
  V(G_1)$  as well as $w \in V(G_2)$ two vertices.
  Then, it holds that
  \begin{align*}
    \tws(G_1 \oplus_{v,w} G_2) & = \max(\tws(G_1), \tws(G_2)) \quad
                               \text{and} \\
        \max(\pws(G_1), \pws(G_2)) \leq \pws(G_1 \oplus_{v,w} G_2) & \leq
    \max(\pws(G_1),\pws(G_2)) + 1.
  \end{align*}
\end{theorem}

\begin{proof}
  The lower bounds follow by Corollary~\ref{cor-pw-tw-subgr}, as $G_1$
  and $G_2$ are subgraphs of $G_1 \oplus_{v,w} G_2$.

  Let $(\calX_1, T_1)$ be a tree-decomposition for $G_1$ of width
  $\tws(G_1)$ and $(\calX_2, T_2)$ be a tree-decomposition for $G_2$ of
  width $\tws(G_2)$.
  To define a tree-decomposition $(\calX, T)$ for $G_1
  \oplus_{v,w} G_2$, we replace every occurrence of $v$ in $\calX_1$ and
  every occurrence of $w$ in $\calX_2$ by $z$.
  Then, we choose a vertex $u_1$ in $V(T_1)$, such that $z$ belongs to
  $X_{u_1} \in \calX_1$, and a vertex $u_2$ in $V(T_2)$, such that
  $z$ belongs to $X_{u_2} \in \calX_2$.
  We define $T$ as the disjoint union of $T_1$ and $T_2$ with the
  additional edge $\{u_1,u_2\}$ and $\calX$ as the union of $\calX_1$
  and $\calX_2$.
  This results in a tree-decomposition $(\calX, T)$ for $G_1
  \oplus_{v,w} G_2$ of width $\max(\tws(G_1),\tws(G_2))$.

  In order to define a path-decomposition for $G_1 \oplus_{v,w} G_2$,
  let $\calX_1$ be a path-decomposition for $G_1$ of width $\pws(G_1)$
  and $\calX_2$ be a path-decomposition for $G_2$ of width
  $\pws(G_2)$.
  Then, we can either proceed as for tree-width, replacing $v$ in all
  bags of $\calX_1$ and $w$ in all bags of $\calX_2$ by $z$, and
  concatenate both paths of bags two a new path $\calX$, resulting in a
  path-decomposition for $G_1 \oplus_{v,w} G_2$ of width
  $\max(\pws(G_1), \pws(G_2))$, or, if the resulting concatenation
  violates (pw-3), additionally add $z$ to all remaining bags of
  $\calX$, resulting in a path-decomposition of width at most
  $\max(\pws(G_1), \pws(G_2)) + 1$.
\end{proof}

If $v \in V(G_1)$ and $w \in V(G_2)$ have degree at least one in $G_1$
and $G_2$, i.e., are no isolated vertices, the new vertex $z$ in $G_1
\oplus_{v,w} G_2$ is called an {\em articulation vertex} of $G_1
\oplus_{v,w} G_2$, since $(G_1 \oplus_{v,w} G_2) - z$ has more
connected components than $G_1 \oplus_{v,w} G_2$.
For a graph $G$, a maximal, biconnected subgraph without any articulation
vertex is called a {\em block} or {\em biconnected component} of $G$.
The bounds of Theorem~\ref{T1s} for tree-width imply that the
tree-width of a graph equals the maximum tree-width of its biconnected
components.

\begin{corollary}\label{cor-bissc-tw}
  Let $G$ be a graph.
  It holds that the tree-width of $G$ equals the maximum tree-width of
  its biconnected components.
\end{corollary}

Contrarily, the subsequent example shows that
Corollary~\ref{cor-bissc-tw} does not hold for path-width.

\begin{example}
  Denote the vertices of $P_3$ by $V(P_3) = \{ v_1, v_2, v_3 \}$ and
  the vertices of $P_5$ by $V(P_5) = \{ u_1, u_2, u_3, u_4, u_5 \}$.
  Then, the incidence graph $I(K_{1,3})$ from Figure~\ref{F-co} can be
  created as $P_3 \oplus_{v_3, u_3} P_5$.
  We know that $\pws(P_3) = \pws(P_5) = 1$ holds and showed in
  Example~\ref{ex-pw} that $\pws(I(K_{1,3})) = 2$ is true.
  As all biconnected components of $I(K_{1,3})$ are subgraphs of $P_3$
  or $P_5$, it follows that all biconnected components have a path-width
  of one.
  Consequently, the path-width of $I(K_{1,3})$ cannot equal the
  maximum path-width of any of its biconnected components.
\end{example}

However, the bounds of Theorem~\ref{T1s} for path-width imply that the
path-width of a graph can be bounded by its number of biconnected
components and their maximum path-widths.

%%%%%%%%%%%%%%%%%%%%%%%%%%%%%%%%%%%%%%%%%%%%%%%%%%%%%%%%%%%%%%%%%%%%%%%%%%%
\subsection{Corona}\label{sec-corona}
Frucht and Harary~\cite{FH70} introduced the corona of two graphs when they
constructed a graph whose automorphism group is the wreath
product of the two graphs' automorphism groups.\footnote{Please be
  unconcerned, the corona of graphs has nothing to do with the global
  pandemic of coronavirus disease 2019 (COVID-19).}
The {\em corona} of two vertex-disjoint graphs $G_1$ and $G_2$,
denoted by $G_1 \wedge G_2$, consists of the disjoint union of one
copy of $G_1$ and $|V(G_1)|$ copies of $G_2$, where each vertex of
the copy of $G_1$ is connected to all vertices of one copy of $G_2$.
In other words, $|V(G_1)||V(G_2)|$ edges are inserted in the
disjoint union of the $|V(G_1)| + 1$ graphs.

Alternatively, the corona of $G_1$ and $G_2$ can also be obtained by
applying 1-sum and dominating vertex operations as follows.
Let $V(G_1) = \{v_1, \ldots, v_n\}$ be the vertex set of $G_1$.
For $i = 1, \ldots, n$, we take a copy of $G_2$, insert a dominating
vertex $w_i$ (cf.\ Section \ref{vertexadd}) in that copy, and obtain
the resulting graph $G_{2,i}$.
Then, the subsequent sequence of 1-sums,
\begin{align}
  \left(\ldots((G_1 \oplus_{v_1,w_1} G_{2,1}) \oplus_{v_2,w_2}
  G_{2,2}) \ldots \right) \oplus_{v_n,w_n} G_{2,n}, \label{1sum}
\end{align}
results in the corona $G_1 \wedge G_2$ of $G_1$ and $G_2$.

With this observation at hand, we can bound the tree-width
(path-width) of $G_1 \wedge G_2$ in the tree-width (path-width) of its
combined graphs as follows.

\begin{theorem}\label{Tcor}
  Let $G_1$ and $G_2$ be two vertex-disjoint graphs.
  Then, it holds that
  \begin{align*}
    \max(\tws(G_1), \tws(G_2)) & \leq \tws(G_1 \wedge G_2) \leq
                                 \max(\tws(G_1), \tws(G_2)) + 1 \quad \text{and} \\
    \max(\pws(G_1), \pws(G_2)) & \leq \pws(G_1 \wedge G_2) \leq
                                 \max(\pws(G_1),\pws(G_2)) + |V(G_1)|.
  \end{align*}
\end{theorem}

\begin{proof}
  The lower bounds follow by Corollary~\ref{cor-pw-tw-subgr} since $G_1$
  and $G_2$ are subgraphs of $G_1 \wedge G_2$.

  For the upper bounds we make use of our earlier observation that we
  can obtain the corona of $G_1$ and $G_2$ by applying 1-sum and
  dominating vertex operations as described in Equation~(\ref{1sum}).
  By Theorem~\ref{Tvertex} it follows that $\tws(G_{2,i}) \leq
  \tws(G_2) + 1$ and $\pws(G_{2,i}) \leq \pws(G_2) + 1$.
  By Theorem~\ref{T1s} and Equation~(\ref{1sum}) it follows that
  $\tws(G_1 \wedge G_2) \leq \max(\tws(G_1), \tws(G_2)) + 1$ and
  $\pws(G_1 \wedge G_2) \leq \max(\pws(G_1),\pws(G_2)) + |V(G_1)|$.
\end{proof}

With the previous theorem proved, we asked ourselves whether there
exists a constant integer $c \in \N$, such that for all graphs $G_1,
G_2$ it holds that $\pws(G_1 \wedge G_2) \leq \max(\pws(G_1),
\pws(G_2)) + c$, similar to the upper bound for tree-width.
The subsequent proposition provides a negative answer to this
question.

\begin{proposition}
  For $n \geq 2$ and $m \geq 1$ it holds that
  \begin{align*}
    \pws(K_n \wedge K_m) = n + \max \left( 0, m - \left\lfloor
    \frac{n}{2} \right\rfloor \right) - 1.
  \end{align*}
\end{proposition}

\begin{proof}
  We write $K_n = (\{ v_1, \ldots, v_n \}, \{ \{ v_i, v_j \} \mid 1
  \leq i < j \leq n \})$ as well as $K_m = (\{ u_1, \ldots, u_m \}, \{
  \{ u_i, u_j \} \mid 1 \leq i < j \leq m \})$.
  For $1 \leq i \leq n$ we
  denote the $n$ copies of $K_m$ in $K_n \wedge K_m$ as $K_m^i$ with
  $V(K_m^i) = \{ u_1^i, \ldots, u_m^i \}$.
  By construction of $K_n
  \wedge K_m$, every vertex $v_i$ of $K_n$ gets connected to all
  vertices of $K_m^i$, resulting in $n$ cliques of size $m + 1$ which
  we denote by $K_{m+1}^i$ with $V(K_{m+1}^i) = V(K_m^i) \cup \{ v_i \}$
  for $1 \leq i \leq n$.

  Since $K_{m+1}^i$, $1 \leq i \leq n$, and $K_n$ are subgraphs of
  $K_n \wedge K_m$, it follows by Corollary~\ref{cor-pw-tw-subgr} that
  $\max(n - 1, m)\leq \pws(K_n \wedge K_m)$ must hold.

  Next, let us construct the ensuing path-decomposition $\calX =
  (X_1, \ldots, X_{n+1})$ for $K_n \wedge K_m$.
  We define
  \begin{enumerate}
  \item
    $X_i = V(K_m^i) \cup \{ v_1, \ldots, v_i \}$ for $1 \leq
    i \leq \left\lfloor \frac{n}{2} \right\rfloor$,
  \item
    $X_{\left \lfloor \frac{n}{2} \right
    \rfloor + 1} = V(K_n)$, and
  \item
    $X_{i + 1} = V(K_m^{i}) \cup \{ v_i, \ldots, v_n \}$ for
    $\left \lfloor \frac{n}{2} \right \rfloor < i \leq n$.
  \end{enumerate}
  It is easy to check that $\calX$ satisfies all three requirements of
  a path-decomposition for $K_n \wedge K_m$.
  Furthermore, we note that (a) $\max_{1 \leq i \leq \left\lfloor
      \frac{n}{2} \right\rfloor} |X_i| = m + \left\lfloor \frac{n}{2}
  \right\rfloor$, (b) $|X_{\left\lfloor \frac{n}{2} \right\rfloor + 1}| = n$,
  and (c) $\max_{\left\lfloor \frac{n}{2} \right\rfloor < i \leq n}
  |X_{i+1}| = m + n - \left\lfloor \frac{n}{2} \right\rfloor$ holds.

  Next, let us differentiate two cases, $m \leq \left\lfloor
    \frac{n}{2} \right\rfloor$ and $m > \left\lfloor \frac{n}{2}
  \right\rfloor$:
  \begin{description}
  \item[Case~1:] $m \leq \left\lfloor \frac{n}{2} \right\rfloor$

    In this case we have for (a) $m + \left\lfloor \frac{n}{2}
    \right\rfloor \leq \left\lfloor \frac{n}{2}
    \right\rfloor + \left\lfloor \frac{n}{2}
    \right\rfloor \leq n$ and for (c) $m + (n - \left\lfloor
      \frac{n}{2} \right\rfloor) \leq \left\lfloor
      \frac{n}{2} \right\rfloor + n - \left\lfloor
      \frac{n}{2} \right\rfloor = n$, such that
    $\max_{1 \leq i \leq n+1} |X_i| = n$ holds.
    Thus, we know that the path-decomposition $\calX$ has a width of
    $n-1$, yielding
    \begin{align*}
      \max(n - 1, m) = n - 1 \leq \pws(K_n \wedge K_m) \leq n
      - 1.
    \end{align*}
    Consequently, in this case we obtain
    \begin{align*}
          \pws(K_n \wedge K_m) = n - 1 = n + \max\left(0, m -
      \left\lfloor \frac{n}{2} \right\rfloor\right) - 1.
    \end{align*}
  \item[Case~2:] $m > \left\lfloor \frac{n}{2} \right\rfloor$

    In this case we have for (a) $m + n - \left\lfloor \frac{n}{2}
    \right\rfloor = m + \left\lceil \frac{n}{2} \right\rceil \geq m +
    \left\lfloor \frac{n}{2} \right\rfloor$ and for (c) $m + n -
    \left\lfloor \frac{n}{2} \right\rfloor = m + \left\lceil \frac{n}{2}
    \right\rceil \geq \left\lfloor \frac{n}{2} \right\rfloor + 1 +
    \left\lceil \frac{n}{2} \right\rceil = n + 1 \geq n$, such that
    $\max_{1 \leq i \leq n + 1} |X_i| = m + n - \left\lfloor \frac{n}{2}
    \right\rfloor$ holds.
    Thus, the width of path-decomposition $\calX$ is $m + n -
    \left\lfloor \frac{n}{2} \right\rfloor - 1 = n + \max(0, m -
    \left\lfloor \frac{n}{2} \right\rfloor) - 1$ and it follows that
    \begin{align*}
      \pws(K_n \wedge K_m) \leq n + \max(0, m - \left\lfloor \frac{n}{2}
    \right\rfloor) - 1
    \end{align*}
    holds.
    Lastly, we show that there cannot exist any path-decomposition
    $\calX'$ for $K_n \wedge K_m$ of width smaller than $n + \max(0, m -
    \left\lfloor \frac{n}{2} \right\rfloor) - 1$.
    To do so, let us assume that $\calX'$ is a minimum
    path-decomposition for $K_n \wedge K_m$ of smallest possible width.

    As $K_n$ is a subgraph of $K_n \wedge K_m$, we know by
    Lemma~\ref{le-pw-clique} that there must exist at least one bag $X \in
    \calX'$ with $V(K_n) \subseteq X$.
    Furthermore, as $K_{m+1}^i$ is a subgraph of $K_n \wedge K_m$ for
    $1 \leq i \leq n$, we know by the same argument that there must exist
    at least one bag $Y_i \in \calX'$ with $V(K_{m+1}^i) \subseteq Y_i$
    for every $1 \leq i \leq n$.

    Assuming that $X$ and $Y_i$, $1 \leq i \leq n$, contain no more
    than the previously mentioned sets of vertices,
    $V(K_n \wedge K_m) = X \cup \bigcup_{i=1}^n Y_i$ is true and for every
    edge $\{u,v\} \in E(K_n \wedge K_m)$ it holds that $u,v \in X$ if $u,v
    \in V(K_n)$ or $u,v \in Y_i$ if $u,v \in
    V(K_{m+1}^i)$.
    Consequently $\calX'$ satisfies (pw-1) and (pw-2) already.

    Thus, it is left to argue that (i) for every $1 \leq i \leq n$ the
    only bag containing $V(K_m^i)$ is $Y_i$ and (ii) the least
    bag-size increasing way to satisfy (pw-3) for $X, Y_1, \ldots,
    Y_n$ is by introducing additional copies of subsets of $V(K_n)$ to
    $Y_1, \ldots, Y_n$.

    For every $u_j^i \in V(K_m^i)$, $1 \leq j \leq m$, $1 \leq i \leq
    n$, it holds that $N_{K_n \wedge K_m}(u_j^i) \subseteq
    Y_i$.
    Consequently, if $u_j^i$ is present in any other bag of $\calX'$
    except $Y_i$, this contradicts our assumption of $\calX'$ being
    minimal, as we could remove $u_j^i$ from all bags in $\calX'$ except
    $Y_i$ and still have a valid path-decomposition.
    Hence, (i) holds.

    Since $\calX'$ must form a path, there can exist at most one bag
    preceding $X$ and one bag succeeding $X$ in the path.
    The most efficient way to ensure that (pw-3) is satisfied is by
    allocating the first $\left\lfloor \frac{n}{2} \right\rfloor$ vertices
    from $X$ to its predecessor and the remaining $\left\lceil \frac{n}{2}
    \right\rceil$ vertices to its successor.
    Thus, (ii) holds and $\calX' = \calX$ follows.

    Consequently, $\calX'$ has a width of at least $n + \max(0, m -
    \left\lfloor \frac{n}{2} \right\rfloor) - 1$, as well, such that
    \begin{align*}
      \pws(K_n \wedge K_m) = n + \max(0, m -
    \left\lfloor \frac{n}{2} \right\rfloor) - 1
    \end{align*}
    follows.
  \end{description}
This shows the statement of the proposition.  
\end{proof}

Note that for $n = 1$ one can verify that $\pws(K_1 \wedge K_m) = \pws(K_{m+1})
= m$ holds.

%%%%%%%%%%%%%%%%%%%%%%%%%%%%%%%%%%%%%%%%%%%%%%%%%%%%%%%%%%%%%%%%%%%%%%%%%%%
%%%%%%%%%%%%%%%%%%%%%%%%%%%%%%%%%%%%%%%%%%%%%%%%%%%%%%%%%%%%%%%%%%%%%%%%%%%
%%%%%%%%%%%%%%%%%%%%%%%%%%%%%%%%%%%%%%%%%%%%%%%%%%%%%%%%%%%%%%%%%%%%%%%%%%%
\section{Conclusions and Outlook}\label{sec-con}

In Section~\ref{sec-un}, we have shown how the tree-width or
path-width of a given graph changes, if we apply a certain, unary
graph transformation $f$ to this graph.
In all cases, in which it is possible to bound the tree-width or
path-width of the resulting graph $f(G)$, we have also shown how to
compute the corresponding decomposition in time linear in the size of
the corresponding decomposition for $G$.
Table~\ref{Ta-unary} summarizes the results.
It is noteworthy that the behavior of tree-width and path-width under
the considered transformations is almost identical.

\begin{table}[ht]
  \centering
  \begin{tabular}{lll}
    \toprule
    transformation $f$     & $\tws(f(G))$  &  $\pws(f(G))$  \\
    \midrule
    vertex deletion        & $k$     & $k$ \\
    vertex addition        & $k+1$   & $k+1$ \\
    edge deletion          & $k$     & $k$ \\
    edge addition          & $k+1$   & $k+1$ \\
    subgraph               & $k$        &  $k$ \\
    vertex identification  & $k+1$   & $k+1$ \\
    edge contraction       & $k$        & $k$ \\
    edge subdivision       & $k$        & $k+1$ \\
    minor                  &  $k$    &  $k$ \\
    switching              &  $k+1$       & $k+1$ \\
    \bottomrule
  \end{tabular}
  \caption{Let $G$ be a graph of tree-width (path-width) $k$ and $f$ the
    unary graph transformation in the first column.
    The second column lists the upper bound for the tree-width and the
    third column the upper bound for the path-width of $f(G)$.\label{Ta-unary}}
\end{table}

Furthermore, in Section~\ref{section-binary-operations}, we have
considered various binary graph operations $f$, which create a new
graph $f(G_1,G_2)$ out of two graphs $G_1$ and $G_2$.
In all cases, in which it is possible to bound the tree-width or
path-width of the combined graph $f(G_1,G_2)$ in the tree-width or
path-width of $G_1$ and $G_2$, we have shown how to compute the
corresponding decomposition in time linear in the size of the
corresponding decompositions for $G_1$ and $G_2$, such that our
results are constructive.
In Table~\ref{Ta-binary}, we summarize these results, which show that, with
an exception for the corona of two graphs, the behavior of tree-width
and path-width under the considered operations is nearly identical.

\begin{table}[ht]
  \centering
  \begin{tabular}{lll}
    \toprule
    operation $f$          & $\tws(f(G_1, G_2))$             & $\pws(f(G_1, G_2))$ \\
    \midrule
    disjoint union         & $\max(k_1, k_2)$                & $\max(k_1, k_2)$ \\
    join                   & $\min(k_1+ n_2, k_2+ n_1)$      & $\min(k_1+ n_2, k_2+ n_1)$ \\
    substitution           &   $\min(k_1+ n_2, k_2+ n_1) -1$ & $\min(k_1 + n_2, k_2+ n_1) -1$ \\
    lexicographic product  & $(k_1+1)  n_2 - 1$              & $(k_1+1)  n_2 - 1$ \\
    1-sum                  & $\max(k_1, k_2)$                & $\max(k_1, k_2)+1$ \\
    corona                 & $\max(k_1, k_2)+1$              & $\max(k_1, k_2) +  n_1$ \\
    \bottomrule
  \end{tabular}
  \caption{Let $G_1$ and $G_2$ be two graphs of tree-width
    (path-width) $k_1$ and $k_2$, respectively, and $f$ the binary graph
    operation from the first column.
    By $n_1$ we denote the number of vertices of $G_1$ and by $n_2$
    the number of vertices of $G_2$.
    The second column lists the upper bound of the
    tree-width and the third column lists the
    upper bound of the path-width for graph $f(G_1,G_2)$.\label{Ta-binary}}
\end{table}

The results in Sections~\ref{edgeadd} and~\ref{sec-sg} allow to
generalize known results on the stability of trees and
forests \cite{WR21,WR21a} to the stability of graph classes of
bounded tree-width~\cite{GRW24}.

Most of our results provide tight upper and lower bounds for the
tree-width and path-width of the resulting graph in terms of the
tree-width and path-width of the initial graphs or argue why such
bounds are impossible.

For the following case it remains to show that our bounds are
the best possible ones or to provide stricter bounds.
The bounds for the tree-width and path-width of the power of a graph,
as shown in Section~\ref{section-power}, are very rough as all vertex
degrees are approximated by the maximum degree of the graph.
Furthermore, neither did we provide lower bounds for the tree-width
(path-width) of the categorical or co-normal graph product of two
graphs, nor for the symmetric difference or rejection of two graphs.

%%%%%%%%%%%%%%%%%%%%%%%%%%%%%%%%%%%%%%%%%%%%%%%%%%%%%%%%%%%%%%%%%%%%%%%%%%%

\end{document}